\documentclass[aps,onecolumn,11pt,floatfix,altaffilletter,preprintnumbers,
               tightenlines,showpacs,showkeys,notitlepage,nofootinbib]{revtex4-1}

\usepackage[colorlinks=true,citecolor=blue,linkcolor=blue]{hyperref}
\usepackage[normalem]{ulem}
\usepackage{amsmath,amssymb}
\usepackage{epsfig}
\usepackage{slashed}
\usepackage{graphicx}               
\usepackage{url}
\usepackage{color}
\usepackage{multirow}
\usepackage{placeins}
\usepackage[dvipsnames]{xcolor}
\usepackage{epstopdf}
\usepackage{tikz}
\usetikzlibrary{trees}
\usetikzlibrary{decorations.pathmorphing}
\usetikzlibrary{decorations.markings}
\usepackage{bbold}

\definecolor{jblue}  {RGB}{20,50,100}
\definecolor{npurple}  {RGB} {153, 51, 204}
\definecolor{wred}   {RGB}{217,0,56}
\definecolor{white}   {RGB}{255,255,255}

\definecolor{korange}   {RGB}{235, 80,  43}
\definecolor{korange2}   {RGB}{245, 100,  63}
\definecolor{kyelloworange}   {RGB}{255, 210,  110}
\definecolor{kyelloworange2}   {RGB}{240, 170,  90}
\definecolor{kred}   {RGB}{204,  102, 153}
\definecolor{kpurple}   {RGB}{153,  61, 190}
\definecolor{kpurplelight}   {RGB}{213,  161, 230}

\usepackage{cleveref}
\usepackage{subfigure} 
\usepackage{lipsum}
\definecolor{red}{rgb}{1.0, 0, 0}
 \usepackage{gensymb}
\allowdisplaybreaks

\bibpunct{[}{]}{,}{n}{}{,}

\newcommand{\diag}{\text{diag}}

\newcommand{\keV}{\,\mathrm{keV}}
\newcommand{\GeV}{\,\mathrm{GeV}}
\newcommand{\TeV}{\,\mathrm{TeV}}
\newcommand{\dd}{\mathrm{d}}

\pacs{}
\keywords{}

\begin{document}

\title{The New $\nu$MSM ($\nu\nu$MSM) : \\ \vspace{0.2 cm} Radiative Neutrino Masses, keV-Scale Dark Matter and Viable Leptogenesis with sub-TeV New Physics } 
         
\author{Sven Baumholzer$^{1}$}            \email{sbaumhol@students.uni-mainz.de}
\author{Vedran Brdar$^{2}$}	       \email{vbrdar@mpi-hd.mpg.de}
\author{Pedro Schwaller$^{1}$\,}            \email{pedro.schwaller@uni-mainz.de}
\affiliation{$^1$PRISMA Cluster of Excellence and
             Mainz Institute for Theoretical Physics,
            Johannes Gutenberg-Universit\"{a}t Mainz, 55099 Mainz, Germany\\
$^2$Max-Planck-Institut f\"ur Kernphysik, Saupfercheckweg 1,
          69117 Heidelberg, Germany            }

\preprint{MITP/18-050}

\begin{abstract}
We study the scenario in which the Standard model is augmented by three generations of right-handed neutrinos and a scalar doublet. The newly introduced fields share an odd charge under a $\mathbb{Z}_2$ parity symmetry. This model, commonly known as ``Scotogenic", was designed to provide a mechanism for active neutrino mass generation as well as a viable dark matter candidate. In this paper we consider a scenario in which the dark matter particle is at the keV-scale. Such particle is free from X-ray limits due to the unbroken parity symmetry that forbids the mixing between active and right-handed neutrinos. The active neutrino masses are radiatively generated from the new scalars and  the two heavier right-handed states with $\sim \mathcal{O}(100)$ GeV masses. These heavy fermions can produce the observed baryon asymmetry of the Universe through the combination of  Akhmedov-Rubakov-Smirnov mechanism and recently proposed scalar decays. To the best of our knowledge, this is the first time that these two mechanisms are shown to be successful in any radiative model. We identify the parameter space where the successful leptogenesis is compatible with the observed  abundance of dark matter as well as the measurements from  the neutrino oscillation experiments. Interestingly, combining dark matter production and successful leptogenesis gives rise to strict limits from big bang nucleosynthesis which do not allow the  mass of dark matter to lie above $\sim 10$ keV, providing a phenomenological hint for considered low-scale dark matter. By featuring the keV-scale dark matter free from stringent X-ray limits, successful baryon asymmetry generation and non-zero active neutrino masses, the model is a direct analogue to the $\nu$MSM model proposed by Asaka, Blanchet and Shaposhnikov. Therefore we dub the presented framework as ``The new $\nu$MSM" abbreviated as $\nu\nu$MSM. 
\end{abstract}

\maketitle

\section{Introduction}
\label{sec:intro}
The Standard model (SM) is a remarkably accurate theory. Its particle content and 
interactions are almost flawlessly mapping Nature's choice. However, there are still several open questions and in particular those that stand out are:
\begin{itemize}
\item{What is the origin of neutrino mass?}
\item{Does dark matter (DM) interact with SM particles?}
\item{Through which mechanism was the observed asymmetry between matter and antimatter generated?}
\end{itemize}

The indisputable answer to any  of these questions would indicate a tremendous progress for physics, in particular for the community striving to discover ``new physics", as it is by now clear that the solution does not lie within the SM.

We attempt to address all of the above questions within the model proposed by Ma \cite{Ma:2006km}, dubbed ``Scotogenic", in which the SM field content is supplemented by three right-handed neutrinos and a scalar doublet, all odd under a postulated $\mathbb{Z}_2$ parity symmetry. The model was originally envisioned to account for small active neutrino masses and the  electroweak scale DM produced via  standard thermal freeze-out. Instead, we put in focus keV-scale for the mass of the fermionic DM particle.       
We consider two viable regimes for DM production and ensure that the studied parameter space is fully consistent with leptonic mixing parameters and the observed  neutrino mass squared differences. 

In addition, we scrutinize the generation of BAU (baryon asymmetry of the Universe) from the produced lepton number asymmetry (leptogenesis). Such  asymmetry transfer is viable due to the existence of  non-perturbative sphaleron processes at high temperatures. 
 We consider two complementary mechanisms, namely BAU production from right-handed neutrino oscillations introduced by Akhmedov, Rubakov and Smirnov (ARS) \cite{Akhmedov:1998qx} as well as recently proposed asymmetry generation from scalar decays \cite{Hambye:2016sby}.

The motivation for this work stems from the $\nu$MSM model proposed by Asaka, Blanchet and Shaposhnikov \cite{Asaka:2005an,Asaka:2005pn,Shaposhnikov:2008pf}. In the $\nu$MSM, the SM particle content is extended with only three right-handed neutrinos, where the lightest one is a keV-scale DM  produced via neutrino oscillations \cite{Shi:1998km,Dodelson:1993je} due to the mixing between active and right-handed neutrinos. The heavier two GeV-scale right-handed states generate  active neutrino masses in the seesaw type-I model \cite{Goran,Minkowski,GellMann:1980vs,Yanagida:1979as} as well as produce BAU through the ARS mechanism. When confronted with the current experimental data, the $\nu$MSM is seriously challenged. In particular, the vast portion of the viable parameter space for DM  is  excluded  by the combination of structure formation and X-ray limits \cite{Baur:2017stq,Perez:2016tcq,Merle:2015vzu}.

A keV-scale DM candidate and ARS mechanism for baryogenesis are  also prominent characteristics in our scenario. Furthermore, the relevant fermionic Yukawa and mass terms in these two models differ only in the employed scalar doublet -- the SM Higgs doublet is considered in $\nu$MSM whereas our model hinges on the existence of a second $\mathbb{Z}_2$-odd scalar doublet. In contrast with $\nu$MSM, in the Scotogenic setup keV-scale DM  does not 
mix with the active sector due to the imposed $\mathbb{Z}_2$ symmetry. Hence, the DM parameter space opens up  due to the absence of astrophysical X-ray limits \cite{Horiuchi:2013noa,Malyshev:2014xqa,Tamura:2014mta, Aharonian:2016gzq}. Let us note that, in such a framework, the controversial 3.5 keV line \cite{Bulbul:2014sua,Boyarsky:2014jta, Cappelluti:2017ywp, Abazajian:2017tcc, Brdar:2017wgy} does not have  a DM origin and the atomic physics explanation is favored \cite{Jeltema:2014qfa, Jeltema:2014mla, Shah:2016efh, Aharonian:2016gzq}.

As in the original $\nu$MSM, we successfully identify the parameter space corresponding to non-zero neutrino masses, correct DM relic abundance and the observed amount of baryon asymmetry in the Universe. Hence, we simultaneously address all three of the aforementioned questions. The 
strongest constraint on our model, arising from the measurements of primordial abundances of light nuclei, forbids the DM candidate to be heavier than $\sim 10$ keV.

The structure of the paper is as follows. In \cref{sec:model} we introduce the model and discuss relevant theoretical and experimental limits. In particular, we show how the neutrino masses are generated and discuss conditions which ensure the correct low-energy neutrino phenomenology (mixing angles and mass squared differences) in our numerical scans. In  \cref{sec:production,sec:leptogenesis} we identify the viable parameter space for DM and BAU in the model, respectively. Results from these two sections are combined in \cref{sec:combined} with the purpose of finding the
regions where DM and BAU can be addressed simultaneously.
In \cref{sec:detection} we discuss the implications from structure formation and feasibility of probing this model at various experimental facilities. Finally, in \cref{sec:summary} we conclude.
  
\section{The Model}
\label{sec:model}

We supplement the Standard Model particle content with an additional scalar doublet 
$\Sigma= (\sigma^{+},\sigma^{0})^T$, and three right-handed neutrinos $N_{i}$.  We study the spectrum in which all new scalars are heavier than the right-handed neutrinos, of which the lightest one, $N_{1}$, is a keV-scale DM candidate.
 We assume that all these newly introduced degrees of freedom
have an odd $(-)$ charge under a $\mathbb{Z}_2$ parity symmetry, whereas the SM particles have an opposite, even $(+)$ charge.
 The scalar sector is  therefore equivalent to the one in the inert doublet model \cite{Barbieri:2006dq} with the potential that reads  \cite{Ma:2006km,Branco:2011iw}

\begin{align}
V=&\mu_{1}^2\, \Phi^\dag \Phi +\mu_{2}^2\, \Sigma^\dag \Sigma+\frac{1}{2}\,\lambda_1\, (\Phi^\dag \Phi)^2+\frac{1}{2}\,\lambda_2 \,(\Sigma^\dag \Sigma)^2 \nonumber \\&+\lambda_3\,(\Phi^\dag \Phi)(\Sigma^\dag \Sigma)+\lambda_4\,(\Phi^\dag \Sigma)(\Sigma^\dag \Phi)+\frac{\lambda_5}{2}\,\left((\Phi^\dag \Sigma)^2+\text{h.c.}\right),
\label{eq:potential}
\end{align}
where $\Phi=(\phi^+,\phi^0)^T$ is the SM Higgs doublet containing the Higgs boson with mass  equal to $2\lambda_1 v^2$, where
 $v=246/\sqrt{2}$ GeV denotes the vacuum expectation value of $\phi^0$.

 By introducing an additional doublet, the scalar sector contains four additional degrees of freedom with masses
\begin{align}
 &m_{\pm}^2 =  \mu_2^2 + \lambda_3 v^2, \nonumber \\ &
  m_S^2 = \mu_2^2+(\lambda_3+\lambda_4+\lambda_5)v^2, \nonumber \\ &
   m_A^2 = \mu_2^2+(\lambda_3+\lambda_4-\lambda_5)v^2.
   \label{eq:ms}
\end{align}
Here, $m_{\pm}$, $m_S$ and $m_A$ are the masses of the charged, CP-even and CP-odd scalar, respectively. In the remainder of the paper we will denote charged scalars with $\sigma^{\pm}$ and neutral CP-even (CP-odd) scalar with $S\,(A)$.

In the fermion sector, 
the presence of a $\mathbb{Z}_2$ symmetry forbids the ``traditional" lepton portal 
\begin{align}
y_\phi \bar{N}\, \tilde{\Phi}^\dag L+\text{h.c.},
\label{eq:Lepton_Sector_forbidden}
\end{align}
where $L$ is the SM lepton doublet and $y_\phi$ is a corresponding Yukawa coupling. However, the Yukawa interaction between $N_{i}$, leptons and $\Sigma$ field is allowed. After adding a Majorana mass term for right-handed neutrinos, the
relevant lepton sector Lagrangian reads
\begin{align}
\mathcal{L}\supset y_{i\alpha}\, \bar{N}_i\, \tilde{\Sigma}^\dag L_\alpha+\frac{1}{2} m_{N_i}
\bar{N_i} N_i^c+\text{h.c.},
\label{eq:Lepton_Sector}
\end{align}
where $\alpha$ denotes the SM lepton generations and $m_{N_i}$ is the mass of $i$-th  right-handed neutrino. Without loss of generality we take the right-handed neutrino mass matrix in the diagonal form. Note that  the replacement $\Sigma\to \Phi$  transforms the Yukawa term in \cref{eq:Lepton_Sector}  into the one given in \cref{eq:Lepton_Sector_forbidden}.

\subsection{Relevant Constraints}
\label{subsec:limits}
 
Before discussing a realization of nonzero neutrino masses  we present the most relevant theoretical and experimental constraints for the considered model.

\begin{itemize}
\item \emph{Big Bang Nucleosynthesis}

The most relevant limit arises from the measurement of primordial abundances of light nuclei \cite{Tytler:2000qf,Pettini,Kawasaki:2017bqm}, known as the big bang nucleosynthesis (BBN). These precisely measured values are sensitive to energy injection into the plasma, for example from the decays of long-lived particles during the BBN epoch. In our model, late decays of $N_2$ to keV-scale DM ($N_1$) could spoil the BBN predictions. The decay rate for such process is \cite{Molinaro:2014lfa}

\begin{align}
 \Gamma(N_2 \to N_1 l_\alpha^\pm l_\beta^\mp)=\frac{m_{N_2}^5}{6144 \pi^3 m_\pm^4}\left(
  |y_{1\alpha}|^2 \, |y_{2\beta}|^2 + |y_{1\beta}|^2 \, |y_{2\alpha}|^2\right),
  \label{eq:three_body_decay}
 \end{align}
where $l_\alpha$ are the charged SM leptons of generation $\alpha$. The BBN limit is relevant chiefly due to the tiny $y_{1\alpha}$ couplings,  necessary to keep DM out of the thermal equilibrium with SM bath. The findings from the recent analysis presented in Ref. \cite{Kawasaki:2017bqm} allow us to infer the  masses and abundances of $N_2$ consistent with respect to BBN constraint. A quantitative discussion of the impact of these constraints on our model parameter space is presented in \cref{sec:combined}.   

\item \emph{Structure Formation}

Due to the absence of mixing between active and right-handed neutrinos, X-ray limits \cite{Perez:2016tcq} on keV-scale DM are non-existent within the presented model which opens up a viable parameter space for light right-handed neutrinos. However, one also needs to take into consideration the parameter space excluded by structure formation bounds (\emph{Lyman}-$\alpha$ forests) and Milky Way satellite counts \cite{Schneider:2016uqi}. Throughout this paper, we will show results consistent with the limits arising from the most conservative scenario where the keV-scale DM is assumed to inherit a thermal distribution function with $\langle p \rangle/T \approx 3.1$. In \cref{sec:detection} we discuss how this spectrum can be made colder. From Refs.  
 \cite{Schneider:2014rda,Schneider:2016uqi,Schneider:2017qdf,Merle:2015vzu, Cherry:2017dwu}  we infer that in order to be in accord with the 
Milky Way satellite count, $m_{N_1}\gtrsim 6\,$ keV is viable. The  existing limits from \emph{Lyman}-$\alpha$ forests are stronger but rather controversial due to effects stemming from inter-galactic medium \cite{Kulkarni:2015fga,Cherry:2017dwu} and therefore we do not adopt them.

\item \emph{Lepton Flavor Violation and Scalar Potential}

The scotogenic model predicts  lepton flavor violation processes \cite{Toma:2013zsa} of the type $l_\alpha \rightarrow l_\beta \gamma$ and $l_\alpha \rightarrow 3 l_\beta$. In \cref{tab:constraints_yuk}  we summarize the bounds given in Ref. \cite{Patrignani:2016xqp}. 

\begin{table}[ht!]
\Large
	\begin{tabular}{c|c}
		LFV process\,\, & \,\,BR upper bound \\
		\hline\hline
		$\mu^+ \rightarrow e^+\,\gamma$ & $4.2\cdot10^{-13}$\\
		$\tau^\pm \rightarrow e^\pm\,\gamma$ & $3.3\cdot10^{-8}$\\
		$\tau^\pm \rightarrow \mu^\pm\,\gamma$ & $4.4\cdot10^{-8}$\\
		$\mu \rightarrow 3e$ & $1.0\cdot10^{-12}$\\
		$\tau \rightarrow 3e$ & $2.7\cdot10^{-8}$\\
		$\tau \rightarrow 3\mu$ & $2.1\cdot10^{-8}$
	\end{tabular}
	\caption{Constraints on LFV processes, taken from \cite{Patrignani:2016xqp}. The left column indicates rare LFV processes and the right one shows the corresponding upper limits on the branching ratios (BR). }
	\label{tab:constraints_yuk}
\end{table}
The quartic couplings in the scalar potential receive 
constraints from the requirement of the vacuum stability \cite{Branco:2011iw,Lindner:2016kqk}  
\begin{align}
&\lambda_3 >-\sqrt{\lambda_1 \lambda_2}, & \lambda_3+\lambda_4-|\lambda_5|> -\sqrt{\lambda_1 \lambda_2},
\label{eq:couplings_constants_boundaries}
\end{align} 
where $\lambda_1>0$ and $\lambda_2>0$. \\

We find the other constraints \cite{Ahriche:2017iar}, arising for instance from the compatibility  with the electroweak precision data (Peskin-Takeuchi parameters) \cite{PT}, to be much weaker with respect to the aforementioned ones.\\
\end{itemize}
\subsection{Active Neutrino Masses}
\label{subsec:active_neutrino_masses}
The Scotogenic model is known as one of the most minimal realizations \cite{Zee,Zee2,Zee3,Babu:1987be,Brdar:2013iea,Kumericki:2017sfc,
Das:2017ski,Appelquist:2002me,Appelquist03} of non-vanishing neutrino masses established at oscillation experiments \cite{Fukuda:1998mi,Kajita}. In order to obtain the  general formula for the neutrino mass matrix in the flavor basis, it is required to calculate the self-energy contributions to the active neutrino propagator, arising from the exchange of both $S$ and $A$.
The Majorana neutrino mass matrix reads \cite{Ma:2006km,Toma:2013zsa}
\begin{align}
 (m_\nu)_{\alpha\beta} = \sum\limits_i \frac{y_{i\alpha}y_{i\beta}m_{N_i}}{16\pi^2} \left[ \frac{m_S^2}{m_S^2-m_{N_i}^2}\mathrm{ln}\left(\frac{m_S^2}{m_{N_i}^2}\right) -
 \frac{m_A^2}{m_A^2-m_{N_i}^2}\mathrm{ln}\left(\frac{m_A^2}{m_{N_i}^2}\right) \right]\equiv \sum_i y_{i\alpha} y_{i\beta}\, \Lambda_i ,
 \label{eq:radiative_masses}
\end{align}
where the summation index $i$ denotes the right-handed neutrino generations. For later convenience, we introduced  $\Lambda$ which abbreviates all the mass matrix components apart from the Yukawa couplings. As we will demonstrate in \cref{sec:production}, in order not to overclose the Universe, the Yukawa couplings of $N_1$ ($y_{1\alpha}$) must be suppressed with respect to the second and third generation ones. Therefore,  $N_1$ does not effectively yield a contribution to \cref{eq:radiative_masses} which consequently sets the lightest active neutrino to be massless. 

For obtaining the Yukawa couplings that are in accord with the active neutrino mass squared differences and mixing parameters, we employ the Casas-Ibarra parametrization \cite{Casas:2001sr}. Phenomenologically, following the above discussion on the relevance of only $N_2$ and $N_3$ states for active neutrino mass generation, the corresponding $2\times 3$ Yukawa submatrix reads 
\begin{align}
 Y_{23}=\left(\sqrt{\Lambda_{23}}\right)^{-1}\,R\,\sqrt{m_\nu^\text{diag}}\,U_\mathrm{PMNS}^\dagger.
 \label{eq:Casas-Ibara-Para}
\end{align}
Here, $\Lambda_{23}=\diag(\Lambda_2,\Lambda_3)$ and
 $R$ is an orthogonal matrix  parametrized  with one complex angle  
\begin{align}
R=
\begin{pmatrix}
0 & \cos (\omega - i \,\xi) & -\sin (\omega - i \,\xi) \\
0 & \sin (\omega - i \,\xi) & \cos (\omega - i \, \xi)
\end{pmatrix},
\label{eq:R}
\end{align}
where $\omega$ and $\xi$ are real parameters. The neutrino mass matrix in the mass basis is either 
\begin{align}
m_\nu^\text{diag}\approx\text{diag}\left(0,\sqrt{m_\mathrm{sol}^2},\sqrt{m_\mathrm{sol}^2+m_\mathrm{atm}^2}\right),
\end{align}
 for normal mass ordering or 
 \begin{align}
m_\nu^\text{diag}\approx\text{diag}\left(0,\sqrt{m_\mathrm{atm}^2},\sqrt{m_\mathrm{sol}^2+m_\mathrm{atm}^2}\right),
\end{align}
 for the inverted one. Here, $m_\mathrm{sol}^2$ and  $m_\mathrm{atm}^2$ are solar and atmospheric mass squared differences \cite{PDG}, respectively. Throughout this work we assume a mass spectrum with normal ordering.

 The leptonic mixing matrix $U_\mathrm{PMNS}$ is parametrized with mixing angles $\theta_{ij}$ and phases $\delta_i$ as \cite{PDG}

\begin{align}
U_{\text{PMNS}}=
\begin{pmatrix}
e^{i \frac{\alpha_1}{2}} c_{12} c_{13} & e^{i \frac{\alpha_2}{2}} c_{13} s_{12} & e^{-i \delta} s_{13}\\
e^{i \frac{\alpha_1}{2}} (- c_{23} s_{12} - e^{i \delta} c_{12} s_{13} s_{23}) &
e^{i \frac{\alpha_2}{2}} ( c_{12} c_{23} - e^{i \delta} s_{12} s_{13} s_{23}) &
c_{13} s_{23} \\
e^{i \frac{\alpha_1}{2}} (-e^{i\delta} c_{12} c_{23} s_{13}+s_{12}s_{23}) & 
 e^{i \frac{\alpha_2}{2}} (-e^{i\delta} c_{23} s_{12} s_{13}-c_{12} s_{23}) & c_{13}c_{23}
\end{pmatrix},
\end{align} 
with abbreviations $c_{ij}\equiv \cos \theta_{ij}$ and 
$s_{ij}\equiv \sin \theta_{ij}$.
Considering the Majorana nature of active neutrinos in this model, there is one Dirac $(\delta)$ and 2 Majorana CP phases $(\alpha_1, \alpha_2)$.\\
\\
Having discussed the construction of the phenomenologically viable Yukawa matrices, we can now estimate the constraints on the parameter space from LFV processes given in \cref{tab:constraints_yuk}.
The most stringent bound is coming from $\mu^+ \rightarrow e^+\,\gamma$ process with the corresponding one-loop contribution approximately given by
\begin{align}
\mathrm{BR}(\mu^+ \rightarrow e^+\,\gamma) \approx 2\cdot 10^{-9}\,y^4,
\label{eq:LFV}
\end{align}
for  $m_N\sim\mathcal{O}(200)$ GeV and $m_\pm\sim\mathcal{O}(1)$ TeV. Here, $y$ denotes the magnitude of the $Y_{23}$ entries.

After comparing \cref{tab:constraints_yuk} and \cref{eq:LFV} we obtain $y\lesssim0.1$. By inserting this  value into \cref{eq:radiative_masses} we can approximately set a lower bound on the quartic coupling
\begin{align}
\lambda_5 \gtrsim 4\cdot10^{-8}.
\end{align}

\section{Dark Matter Production}
\label{sec:production}
 
The main goal of this section is to describe the two mechanisms for generating the abundance of keV-scale DM. First, in \cref{sub:decay}, we discuss DM production via its feeble interactions with  
thermalized neutral  ($S$, $A$) and  charged scalars
($\sigma^\pm$). More precisely, the DM abundance gradually increases from the decays of heavy $\mathbb{Z}_2$-odd scalars, a mechanism known as DM ``freeze-in" \cite{Hall:2009bx} (see  \cite{Calibbi:2018fqf,Biswas:2018aib} for most recent studies). 
We calculate the magnitude of the coupling required to obtain the observed DM energy density in the Universe.
 
Furthermore, DM is produced via decays of heavier right-handed neutrinos which  ``freeze-out" \cite{Lisanti:2016jxe} from the thermal bath. The viability of such an option is explored in \cref{sub:N2_decay}. These two production mechanisms do not exclude one another and, in fact, both can significantly contribute in general scenarios.

\subsection{Production via Scalar Decays}
\label{sub:decay}

It is well-known that the standard thermal ``freeze-out" of  keV-scale DM is not feasible within the considered model\footnote{Alternatives, such as the late time entropy production \cite{Bezrukov:2009th,King:2012wg} would serve as a remedy, but then one needs to further extend the particle content of the model. Hence, we do not pursue such option.}.
Therefore, we are led toward the ``freeze-in" production through which the abundance of non-thermalized $N_1$ is built up from the decays or annihilations of particles that are in the thermal bath. One may easily infer \cite{Molinaro:2014lfa}  that, in this model, the dominant  production arises from the decays of 
 neutral and charged scalars $(A, S, \sigma^\pm)$  which are in the thermal bath due to the strong gauge interactions. The relevant processes are
\begin{align}
A,S\to N_1 \,\nu_\alpha,\hspace{5mm}\sigma^\pm\to N_1 l^\pm_\alpha,
\label{eq:decay}
\end{align}
where  $\nu_\alpha$ correspond to SM neutrinos of generation $\alpha$. 
The general form of the Boltzmann equation for the DM production via ``freeze-in" reads \cite{Hall:2009bx}

\begin{align}
\frac{d n_{N_1}}{d t}+3H\,n_{N_1}=\int d\Pi_{N_1}\, d\Pi_{\Sigma}\,d\Pi_{L}\, (2\pi)^4
\delta^{(4)}(p_{N_1}+p_{L}-p_{\Sigma}) |\mathcal{M}|^{2}_{\Sigma\to N_1 L}\, f_{\Sigma}.
\label{eq:Boltzmann}
\end{align}
Here, $n_{N_1}$ is the DM number density, $H$ is the Hubble parameter, $\Pi_{k}$ abbreviates the phase space factor $\left(\frac{d^3 p_k}{2 E_k (2\pi)^3}\right)$ for particle $k$, $|\mathcal{M}|^{2}_{\Sigma\to N_1 L}$ is the squared 
matrix element for the decay of a scalar into $N_1$ and SM lepton, and $f_{\Sigma}$ is the distribution function of scalar particles. For brevity, we  jointly denote the summation over all scalar decay channels (see  \cref{eq:decay}) with $\Sigma\to N_1 L$. 

The formula given in \cref{eq:Boltzmann} can be simplified to the form  \cite{Hall:2009bx}
\begin{align}
\frac{d n_{N_1}}{d t}+3Hn_{N_1}=\frac{1}{2\pi^2}\,\Gamma_{\Sigma\to N_1 L} m_{\Sigma}^2 \,T \,K_1\left(\frac{m_{\Sigma}}{T}\right),
\end{align}
where  $m_{\Sigma}$ is the mass of a decaying scalar, and $K_1$ is the modified Bessel function of the second kind. 

By relating the number density  of $N_1$ with the corresponding yield $Y=n_{N_1}/s$, where $s$ is the entropy density, as well as using $dT=-H\,T\,dt$ we obtain the following differential equation
\begin{align}
\frac{dY}{dT}=-\frac{1}{H T s} \frac{1}{2\pi^2}\,\Gamma_{\Sigma\to N_1 L} m_{\Sigma}^2 \,T \,K_1\left(\frac{m_{\Sigma}}{T}\right).
\label{eq:yield-diff}
\end{align}
After changing the integration variable from $T$ to $x=m_\pm/T$,
as well as expanding \cref{eq:yield-diff} over all $\Sigma$ components, we reach the expression

\begin{align}
\frac{dY}{dx}=\frac{135\, M_{\text{Pl}}\, |y_1|^2}{1.66\cdot 64\,\pi^5 g_*^{3/2} m_\pm} x^3 \bigg(2\,K_1(x)+r_A^3  \,K_1(r_A \, x) + r_S^3\, K_1(r_S\, x) \bigg),
\label{eq:yield-diff-dx}
\end{align}
where we explicitly inserted formulae for $H$, $s$ and the partial widths of the scalars \cite{Molinaro:2014lfa}. For simplicity, we assumed that $y_{1\alpha}$ coupling is identical for each flavor $\alpha$, therefore denoted $y_1$.
In \cref{eq:yield-diff-dx}, $g_{*}$ is the number of the degrees of freedom in the thermal bath at the time of DM production, $M_\text{Pl}$ is non-reduced Planck mass, and $r_A$ ($r_S$)  denotes the scalar mass ratio  $m_A/m_\pm$ $\left(m_S/m_\pm\right)$.

After an explicit integration where we assume the reheating temperature to be much higher than any other mass scale in our model, we obtain the following expression for the present DM yield
\begin{align}
Y_\text{FI}= \frac{405 M_\text{Pl}\, |y_1|^2}{128\pi^4\cdot1.66\cdot g_*^{3/2} m_\pm}\frac{2 r_A r_S +r_S+r_A}{r_A r_S}.
\label{eq:yield-integrated}
\end{align}
 By taking into account all SM and new degrees of freedom, we have $g_*\approx 114.25$ at $\mathcal{O}(10^2-10^3)$ GeV temperature where the freeze-in occurs.
 
For the purpose of estimating the required order of magnitude for $y_1$, let us further simplify \cref{eq:yield-integrated} by assuming $r_A=r_S=1$ which corresponds to the $\lambda_{4,5}\to 0$ limit.  We arrive at
\begin{align}
Y_\text{FI}\approx 7.82\times 10^{11}\, |y_1|^2 \left(\frac{1\,\text{TeV}}{m_\pm} \right).
\label{eq:yield-num}
\end{align}

By using the well-known  relation  between DM yield and the relic abundance 
\begin{align}
\Omega h^2_\text{FI} =2.742\cdot 10^2 \,\bigg(\frac{m_{N_1}}{\text{keV}}\bigg) \,Y_\text{FI},
\label{eq:relic_abundance}
\end{align}
we finally obtain 
\begin{align}
\Omega h^2_\text{FI} \approx 0.12\, \bigg(\frac{|y_1|}{2.36\cdot10^{-8}}\bigg)^2 \bigg(\frac{m_{N_1}}{1\, \text{keV}} \bigg)\bigg(\frac{1\TeV}{m_{\pm}} \bigg).
\label{eq:relic_abundance2}
\end{align}

\begin{figure}[ht!]
  \includegraphics[width=0.75\textwidth]{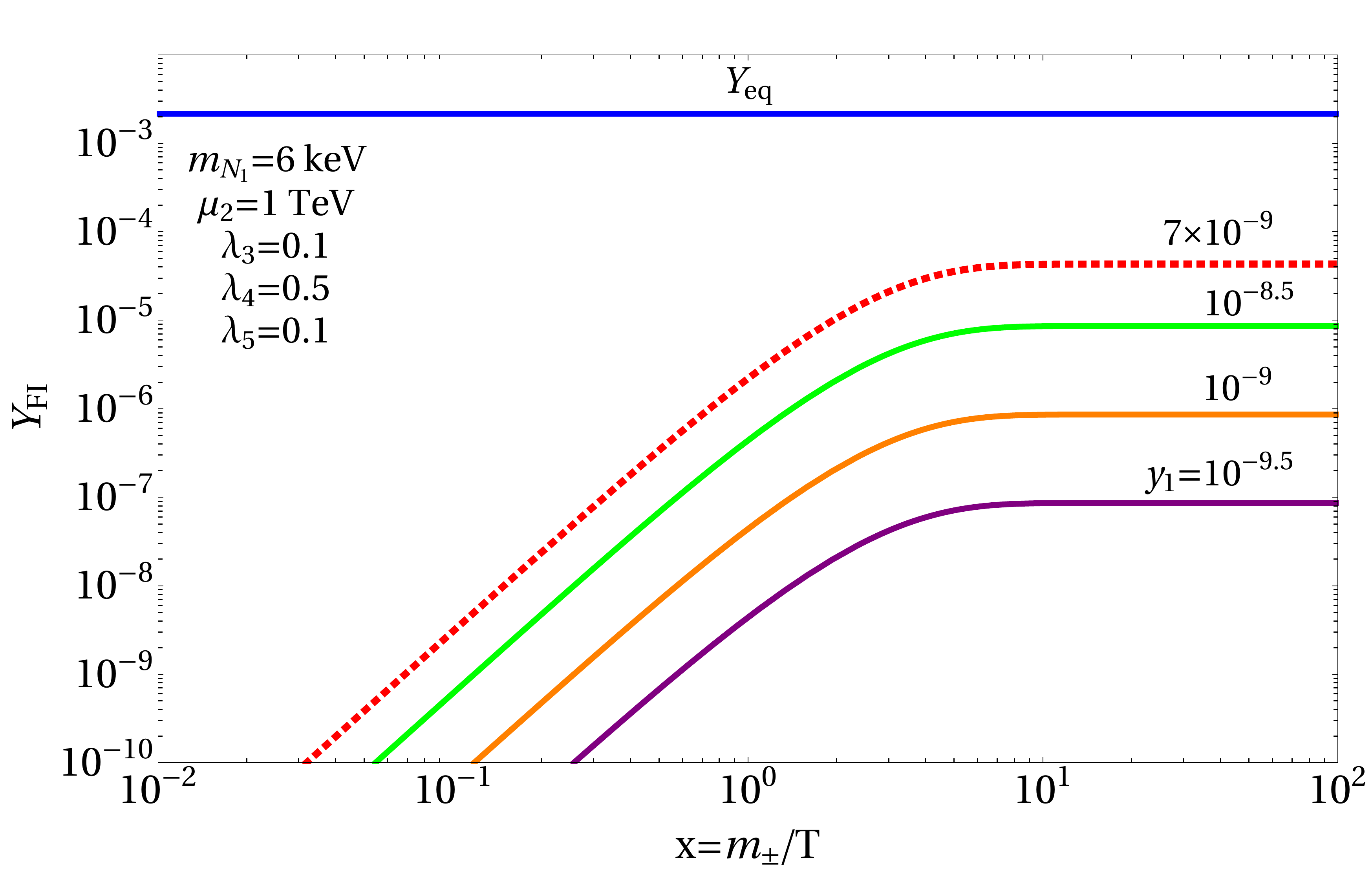}
  \caption{DM freeze-in yield as a function of $x=m_\pm/T$, shown for different values of $y_1$. We have chosen the bare mass parameter $\mu_2 = 1000 \,\text{GeV}$ and fixed the quartic terms to $\lambda_3=\lambda_5=0.1$ and $\lambda_4=0.5$. The parameter choice of the red dashed curve leads to $\Omega h^2_{\mathrm{FI}}=0.12$.}
  \label{figure:plot_yieldNs}
\end{figure}

 In \cref{figure:plot_yieldNs}, we present the numerical solutions of  \cref{eq:yield-diff-dx} for different choices of $y_1$ and fixed $m_{N_1}$. We observe that the dominant production occurs at $\mathcal{O}(1)$ $x$. 
We point out that there is no strong dependence of $Y_\text{FI}$ on the values of quartic couplings. Hence, any physical choice of these parameters (obeying the constraints from \cref{eq:couplings_constants_boundaries}) yields a very similar result to the one presented in \cref{figure:plot_yieldNs}. We show the DM freeze-in for $m_{N_1}=6\keV$, the value in accord with the structure formation limits given in \cite{Merle:2015vzu}. For larger values of DM mass, coupling  $y_1$ needs to be smaller according to \cref{eq:relic_abundance2}.\\  
Let us finally remark that $N_2$ and $N_3$ could in principle also be produced via freeze-in from scalar decays. However, if that was the case, from \cref{eq:relic_abundance2} we infer that the corresponding Yukawa couplings ($y_{2\alpha}$, $y_{3\alpha}$) would then have to be $\ll 10^{-8}$. Such small couplings would not be sufficiently large to generate  eV-scale neutrino masses (see \cref{eq:radiative_masses}). The required strength of ($y_{2\alpha}$, $y_{3\alpha}$) puts $N_2$ and $N_3$ in thermal equilibrium with SM particles.

\subsection{Production via $N_2$ Decays}
\label{sub:N2_decay}
As already discussed in \cref{sec:model}, the $\mathbb{Z}_2$-odd sector consists in total of three right-handed neutrinos,  neutral ($A$ and $S$) and charged scalars $\sigma^\pm$. The mass spectrum is assumed to be $m_{N_1} \ll m_{N_2} \lesssim m_{N_3} < (m_\pm,m_{A},m_{S})$. This spectrum has several appealing features. Since scalars are heavier than  right-handed neutrinos, potential breaking of parity symmetry induced by renormalization group evolution is  evaded \cite{Merle:2015gea}. Additionally, models with a low scale DM, in comparison with conventional $\mathcal{O}(10^2)$ GeV DM, manifest improvement in the predictions for the small scale structure \cite{Ab09}.

Due to small Yukawa couplings of $N_1$ $(y_{1\alpha}\ll y_{2\alpha},y_{3\alpha})$ (see \cref{sub:decay}), the decay rate rate for the process 
$N_2 \rightarrow N_1 L_\alpha \bar{L}_\beta$ is much smaller in comparison to the processes involving heavier $\mathbb{Z}_2$-odd particles. 
 We have numerically checked that such decays occur only after $N_2$ undergoes a successful thermal freeze-out. Therefore, the amount of DM produced from $N_2$ decays is determined by the freeze-out abundance of $N_2$, which we calculate in the following.
 
The Boltzmann equation for each of the considered $\mathbb{Z}_2$-odd  particles ( $N_2,N_3,\sigma^\pm,A,S$, commonly denoted as $\chi_i,\, {i=1...N}$) is given as \cite{Edsjo:1997bg}
\begin{align}
 \frac{d n_i}{d t}+3Hn_i = & -\sum_{j=1}^N \langle \sigma_{ij}v_{ij}\rangle \,(n_in_j-n^\mathrm{eq}_in^\mathrm{eq}_j)\\
 & -\sum_{j\neq 1}^N \left(\langle \sigma^\prime_{Xij} v_{ij}\rangle\,(n_in_X-n^\mathrm{eq}_in^\mathrm{eq}_X) - 
 \langle\sigma^\prime_{Xji} v_{ij}\rangle\,(n_jn_X-n^\mathrm{eq}_jn^\mathrm{eq}_X)\right)\\
 & -\sum_{j\neq i}^N\left(\Gamma_{ij}(n_i-n^\mathrm{eq}_i)-\Gamma_{ji}(n_j-n^\mathrm{eq}_j)\right),
 \label{eq:full-Boltz}
\end{align}
where the terms in the first, second and third row on the right hand side represent (co)annihilations, scatterings and decays, respectively. Here,
 $\sigma_{ij} = \sum\limits_X\sigma(\chi_i\chi_j\rightarrow X)$ is the (co)annihilation cross section for $\mathbb{Z}_2$-odd particles into SM particles, 
$\sigma^\prime_{Xij} = \sum\limits_Y \sigma(\chi_i X \rightarrow \chi_j Y)$ is the cross section for scattering with SM bath (denoted with $X$ and $Y$), and $\Gamma_{ij} =\sum\limits_X \Gamma(\chi_i \rightarrow \chi_j X)$ are decay rates.\\
Since all heavier $\mathbb{Z}_2$ particles eventually decay into $N_2$, we may define the total $N_2$ number density as\footnote{ the procedure also holds for almost degenerate $N_2$ and $N_3$ (see \cref{sec:leptogenesis}).}
\begin{align}
 n=\sum\limits_in_i.
 \label{eq:sum-den}
\end{align}
Considering the size of the coupling constants in our setup, all the $\chi_i$ are in the thermal equilibrium at $\mathcal{O}(1)$ TeV temperatures with the number density equal to 
\begin{align}
 n^{\mathrm{eq}}_i=\frac{T}{2\pi^2}\sum\limits_i g_i \,m_i^2 K_2\left(\frac{m_i}{T}\right),
\end{align}
where $g_i$ denotes the number of internal degrees of freedom of species $i$.
By employing  \cref{eq:full-Boltz,eq:sum-den} and as well as the relation \cite{Edsjo:1997bg} 
\begin{align}
\frac{n_i}{n}\simeq \frac{n_i^\text{eq}}{n^\text{eq}},
\end{align}
we arrive at the standard
Boltzmann equation for the evolution of $N_2$ number density \cite{Belanger:2014hqa}
\begin{align}
 \frac{\dd n}{\dd t} = -3Hn -\langle \sigma_{\mathrm{eff}}v\rangle(n^2-n_\mathrm{eq}^2),
 \label{eq:n-diff-DM}
\end{align}
where $\langle \sigma_{\mathrm{eff}}v\rangle$ is given by \cite{Edsjo:1997bg}
\begin{align}
 \langle \sigma_{\mathrm{eff}}v\rangle = \sum_{i,j} \langle \sigma_{ij} v_{ij} \rangle  \frac{n_{i}^\text{eq}\,n_{j}^\text{eq}}{(n^\text{eq})^2}.
\end{align}
Note that in \cref{eq:n-diff-DM} there are no scattering terms as well as decays. After employing the summation given in \cref{eq:sum-den} these contributions get removed  as they do not change the total number of $\mathbb{Z}_2$-odd particles.

Changing variables to dimensionless quantities $Y=n/s$ and $y=m_{N_2}/T$ yields
\begin{align}
\frac{\dd Y}{\dd y} = \sqrt{\frac{\pi g_*}{45}} \frac{M_\text{Pl}\,m_{N_2}}{y^2}  \langle \sigma_{\mathrm{eff}}v\rangle \left(Y^2_\mathrm{eq}-Y^2\right).
\label{eq:yield_boltzmann}
\end{align}

We have evaluated \cref{eq:yield_boltzmann} with \texttt{micrOMEGAs} \cite{Barducci:2016pcb} and compared with the output of \texttt{MadDM} \cite{Backovic:2013dpa}. 
The Yukawa couplings of $N_2$ and $N_3$ are generated as described in  \cref{subsec:active_neutrino_masses}. Therefore, in the code, only the points in the parameter space that are fully consistent with the neutrino mixing angles and mass squared differences are evaluated.

\begin{figure}[ht!]
	\includegraphics[width=0.85\textwidth]{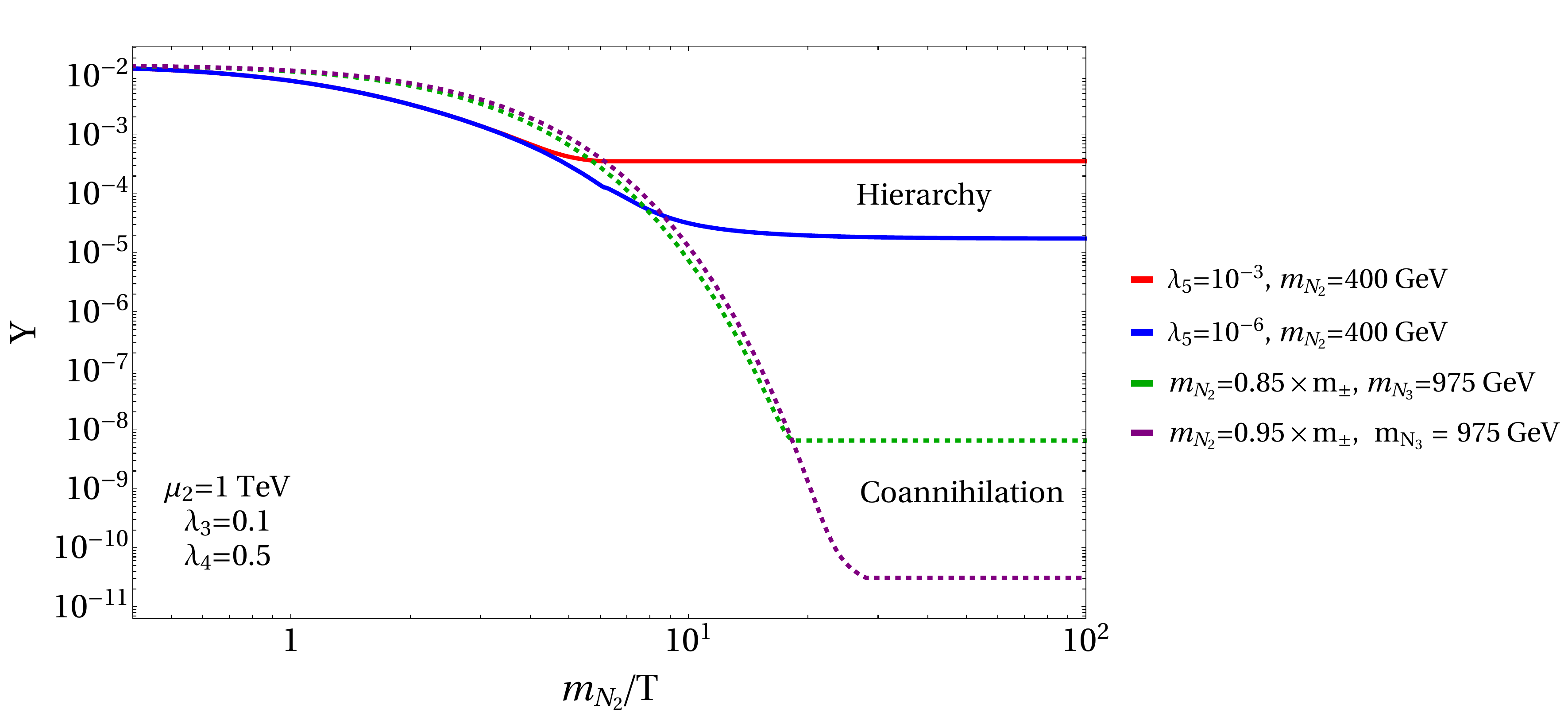}
	\caption{Evolution of the total $N_2$ yield for different values of $\lambda_5$ in a scenario with vastly different scalar and $N_2$ masses (red and blue curve) as well as in the case when the masses are similar  (green and purple dashed curve). In the latter case, due to  larger interaction strength, $N_2$ remains in thermal equilibrium longer, yielding orders of magnitude smaller $N_2$ abundance with respect to the former case.}
	\label{fig:Freeze_Out_N2}
\end{figure}

 In  \cref{fig:Freeze_Out_N2}, we show  the time evolution of $Y$ for several different couplings and right-handed neutrino masses. As can be seen, the yield can reach high values for large $\lambda_5$. This is because, in such case, the Yukawa couplings $y_{2\alpha}$ are small (see \cref{eq:radiative_masses})  and $N_2$ does not remain in thermal equilibrium for long (freeze-out takes place at higher temperatures with respect to  the usual $x\approx20$). 
 
 It is important to point out the coannihilation processes which are dominant if the scalars and $N_{2}$ have similar masses. For $m_\pm>m_{N_2}\gtrsim0.8\,m_\pm$, $Y$ is generally lowered by several orders of magnitude with respect to the general case with hierarchical fermion and scalar masses. This can also be seen in \cref{fig:Freeze_Out_N2} where the green (purple) curve indicates 
 the case with $m_{N_2}=0.85\,m_\pm$\,($m_{N_2}=0.95\,m_\pm$).
 We will particularly emphasize the importance of coannihilations  when exploring viable joint parameter space for DM and leptogenesis (see \cref{sec:combined}). 
 
The contribution to the DM relic abundance from $N_2$ decays is
 
 \begin{align}
 \Omega h^2_{N_2\to N_1}=\frac{m_{N_1}}{m_{N_2}} \,\Omega h^2_{N_2},
 \end{align}
where the relation between $\Omega h^2_{N_2}$ and $Y$ matches the one in \cref{eq:relic_abundance}.

Taking into account the complementary contribution from scalar decays given in \cref{sub:decay}, the requirement for having the amount of DM in accordance with the measurements \cite{Ade:2015xua} is 
\begin{align}
\Omega h^2_{N_2\to N_1} + \Omega h^2_\text{FI}=0.12.
\label{eq:Total_Yield}
\end{align}
As already discussed, we construct the Yukawa couplings of $N_2$ and $N_3$   in agreement with the results from neutrino oscillation experiments. Such a choice already unambiguously fixes $\Omega h^2_{N_2\to N_1}$ contribution, meaning that if it is already overshooting the observed value of $0.12$, the corresponding parameter choice is excluded. 
Otherwise, the Yukawa couplings of $N_1$ could be accommodated in such a way to satisfy \cref{eq:Total_Yield}. Our findings in \cref{sub:decay} indicate $y_{1\alpha}\lesssim 10^{-8}$.

\begin{figure}[ht!]
	\includegraphics[width=0.75\textwidth]{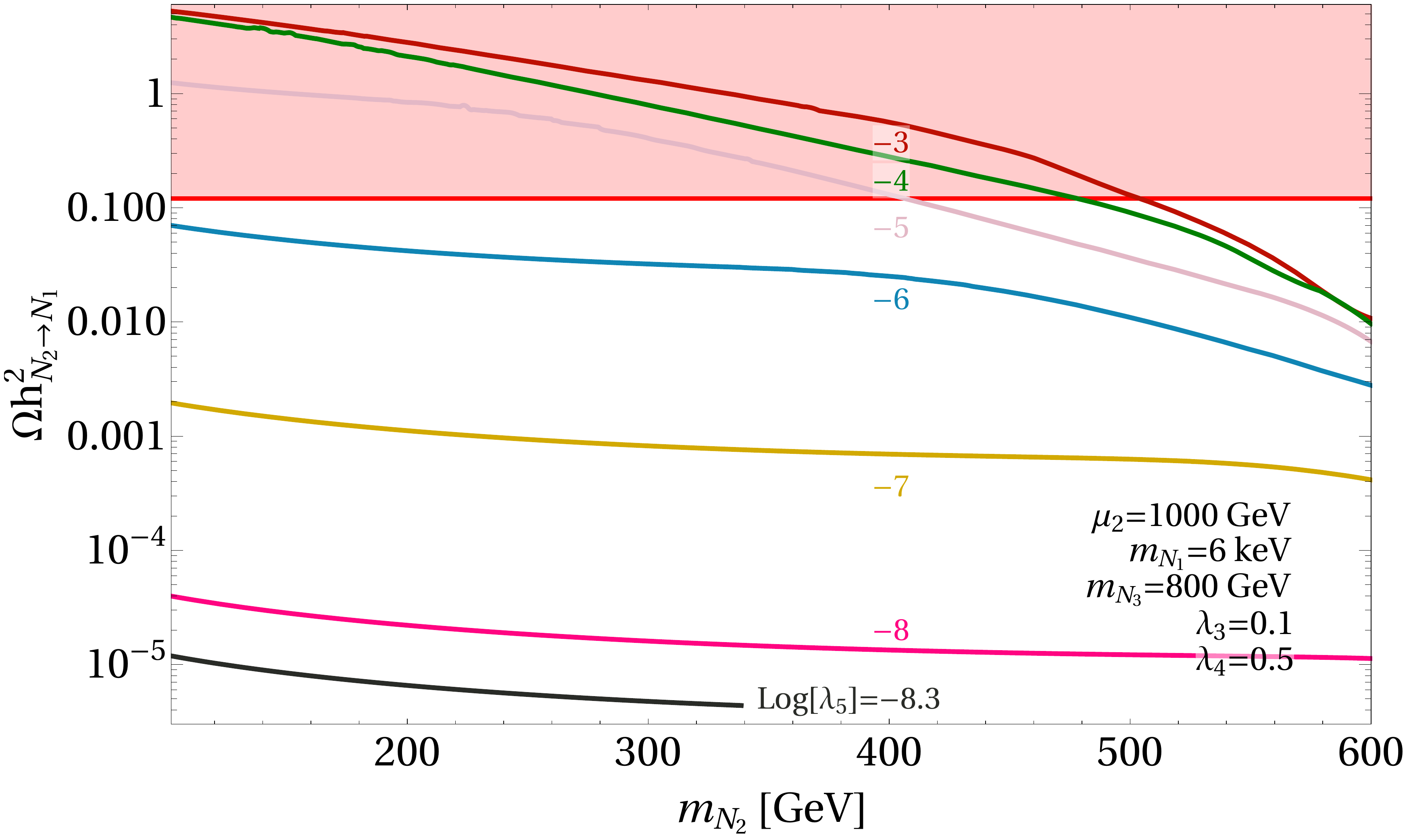}
	\caption{$\Omega h^2_{N_2\to N_1}$ for different choices of the  coupling constant $\lambda_5$. In general, larger values of this coupling lead to larger DM production from $N_2$ decays. Every parameter point in the red shaded region is overproducing DM and is thus ruled out. The truncation of the $\lambda_5=10^{-8.3}$ curve is due to the bounds from LFV processes.}
	\label{fig:N2_Omega_contribution}
\end{figure}

In \cref{fig:N2_Omega_contribution}, results for $\Omega h^2_{N_2\to N_1}$  are shown as a function of $m_{N_2}$. 
For $m_{N_2}\lesssim 500$\, GeV  one may infer  that in the case of larger $\lambda_5$ couplings, $\Omega h^2_{N_2\to N_1}$  poses the dominant contribution to DM relic density since, due to the weakness of $N_2\,N_2\rightarrow \,\text{SM}\,\text{SM}$ channels,  $N_2$ undergoes freeze-out very early.
On the other hand, for higher $m_{N_2}$, additional annihilation channels involving new scalars become dominant, increasing the overall cross section, and correspondingly reducing  $\Omega h^2_{N_2\to N_1}$. In contrast, for very tiny $\lambda_5$ (pink curve), $N_2$ annihilates efficiently even in the absence of coannihilation processes, and therefore the $m_{N_2}$ dependence flattens. We stress here again that $\lambda_5$ can not be pushed to arbitrarily low values, because of the constraints from lepton flavor violation experiments (see \cref{sec:model}). 
 In conclusion, two different behaviors can be seen in the light of \cref{eq:Total_Yield}. For small $\lambda_5$, the parameters $y_{1\alpha}$  and $m_{N_2}$ are practically independent of each other, whereas a cutoff value of $m_{N_2}$ below which DM would be overproduced is reached for larger $\lambda_5$.

\section{Leptogenesis}
\label{sec:leptogenesis}
In order to generate the observed BAU we study the option where initially an asymmetry in the lepton sector is produced \cite{Buchmuller:2004nz}. The production of lepton asymmetry is dubbed leptogenesis. In the simplest realizations, the models featuring leptogenesis incorporate the seesaw type-I mechanism and rely on CP violating decays of hierarchical heavy right-handed neutrinos. The produced lepton asymmetry can be partially converted to the baryon asymmetry due to the existence of non-perturbative processes in thermal equilibrium \cite{KUZMIN198536}. These processes, called sphalerons, violate $B +L$, but conserve $B-L$ numbers and thus allow for an asymmetry conversion between the two sectors. 
When the temperature drops below $T_s = 131.7\GeV$ \cite{DOnofrio:2014rug}, the sphalerons decouple from the thermal bath and the asymmetry conversion ceases.

The drawback of the above mentioned  realization  is that in order to generate the observed BAU the masses of the right-handed neutrinos have to be at least of $\mathcal{O}(10^{8})\GeV$ \cite{Buchmuller:2002rq,Clarke:2015gwa} which is not experimentally reachable\footnote{Taking into account flavor effects~\cite{Barbieri:1999ma,Abada:2006fw,DeSimone:2006nrs,Beneke:2010dz},  recent studies suggest that masses as low as $10^6$~GeV are consistent with successful thermal leptogenesis, which are nevertheless difficult to access experimentally~\cite{Moffat:2018wke}.}. It is, however, possible to lower the needed mass scale  below TeV. In this paper we present such an option in the frame of the Scotogenic model. We rely on two competing mechanisms to create a sufficient baryon asymmetry. 

Firstly, we incorporate oscillations among the right-handed neutrinos (ARS) \cite{Akhmedov:1998qx,Hernandez:2016kel,Drewes:2016gmt,Drewes:2017zyw,Shuve:2014zua} which transfer the asymmetry to the SM leptons
via lepton portal (see \cref{eq:Lepton_Sector}). Secondly, we also take into account decays of the new $\Sigma$ doublet which at finite temperatures serves as an additional source of CP violation. This process is suppressed by a mass insertion factor $(m_{N_2}/T)^2$ and was neglected in the past literature. However, recently it has been shown that scalar decays are important \cite{Hambye:2016sby} and can  even dominate ARS in some regions of  the parameter space \cite{Hambye:2017elz}. Hence, we take both ARS and scalar decays into account.

It is worth noting that several attempts to implement leptogenesis in the Scotogenic model had already been made (see e.g. \cite{Suematsu:2011va,Kashiwase:2012xd}). In Refs.\,\cite{Racker:2013lua,Hugle:2018qbw}, the authors were able to achieve a low scale leptogenesis without imposing mass degeneracies between right-handed neutrinos.\\

For studying leptogenesis, we apply the procedure presented for seesaw type-I mechanism  in Ref. \cite{Hambye:2017elz} to the Scotogenic model. In other words, we replace SM Higgs doublet with $\Sigma$ in the lepton portal term. Only the two right-handed neutrinos, namely  $N_2$ and $N_3$ participate in the lepton asymmetry production.
In the density matrix formalism, the production of a lepton asymmetry via both ARS mechanism and
scalar decays is automatically included. Adopting the notation from \cite{Hambye:2017elz}, we work with yields $\rho$ associated with the number density matrices of  right-handed neutrinos 
\begin{align}
& \rho = \frac{n}{s}, &  \bar{\rho} = \frac{\bar{n}}{s}.
\end{align}
 Here, $n$ is a $2\times 2$ matrix with diagonal elements associated to the densities and the off-diagonal entries parametrizing the mixing between $N_2$ and $N_3$ fields. The states with the opposite helicity are indicated with the overline. In order to calculate the lepton asymmetry  $\delta Y_l$, for each flavor $l$,
we solve the following set of coupled differential equations

\begin{align}
& \frac{d \, \rho_{\alpha \beta}}{d z}  \, = \, - \frac{i}{zH} \, \big[\mathcal E_N,\rho\big]_{\alpha \beta} 
\, - \, \frac{1}{2zHs} \bigg\{\gamma^{LC} \!\! + \! \gamma^{LV}\, , \, \frac{\rho}{\rho^N_{eq}} \! - \! \mathbb{1} \, \bigg\}_{\alpha \beta} \notag\\
& + \, \frac{\delta Y_l}{2 n^L_{eq}zH} \bigg( \!\big( \gamma^{LC}_{WQ,l} \!-\! \gamma^{LV}_{WQ,l}\big)  +  \frac{1}{2} \bigg\{ \gamma^{LC}_{WC,l}\! - \! \gamma^{LV}_{WC,l}\,,\, \frac{\rho}{\rho^N_{eq}}\bigg\}\bigg)_{\alpha \beta}   ,
\label{eq:rate_eq}
\end{align}

\begin{align}
& \frac{d \, \bar \rho_{\alpha \beta}}{d z}  \ = \ - \frac{i}{zH} \, \big[\mathcal E_N, \bar \rho\big]_{\alpha \beta} - \, \frac{1}{2zHs} \bigg\{\, \gamma^{LC \, *} + \gamma^{LV\, *}\,,\, \, \frac{\bar \rho}{\rho^N_{eq}} - \mathbb{1} \, \bigg\}_{\alpha \beta} \notag\\
& - \, \frac{\delta Y_l}{2 n^L_{eq}zH} \bigg( \!\big( \gamma^{LC \,*}_{WQ,l} \! - \! \gamma^{LV\,*}_{WQ,l}\big)  + \frac{1}{2} \bigg\{ \gamma^{LC \, *}_{WC,l} - \gamma^{LV \, *}_{WC,l}\,,\, \frac{\bar \rho}{\rho^N_{eq}}\bigg\} \! \bigg)_{\alpha \beta} \!,
\label{eq:rate_eq_bar}
\end{align}

\begin{align}
\frac{d \, \delta Y_l}{d z}  \ &= \ \frac{1}{zHs\rho_{eq}^N} \text{Tr}\, \bigg\{\big( \gamma^{LC}_l - \gamma^{LV}_l \big) \, \rho \bigg\} \notag\\
&-\frac{1}{zHs\rho^N_{eq}} \text{Tr}\, \bigg\{\big( \gamma^{LC \,*}_l - \gamma^{LV \,*}_l \big) \, \bar \rho \bigg\} - \frac{\delta Y_l}{zH n^L_{eq}} \text{Tr}\, \bigg\{ \gamma^{LC}_{WQ,l} + \gamma^{LV}_{WQ,l}\bigg\} \notag\\& - \frac{\delta Y_l}{2zH n^L_{eq}} \frac{1}{\rho^N_{eq}} \text{Tr}\, \bigg\{\rho (\gamma^{LC}_{WC,l} + \gamma^{LV}_{WC,l}) \bigg\} 
 - \frac{\delta Y_l}{2zH n^L_{eq}} \frac{1}{\rho^N_{eq}} \text{Tr}\, \bigg\{ \bar \rho (\gamma^{LC \,*}_{WC,l} + \gamma^{LV \, *}_{WC,l})\bigg\},
 \label{eq:rate_eq_asym}
\end{align}
where the parameter $z$ equals $z=T_s/T$ and $T_s \sim 131.7\GeV$ is the temperature at which sphalerons decouple. Integration is performed between 
$z_{\text{start}}=T_\mathrm{sph}/T=10^{-3}$ and $z_{\text{stop}}=1$ with the initial conditions 
$\rho_{\alpha\beta}(z_{\text{start}})=0,\,\bar{\rho}_{\alpha\beta}(z_{\text{start}})=0, \,\delta Y_l(z_{\text{start}})=0$, where indices $\alpha$ and $\beta$ run from 1 to 2 due to the relevance of only two right-handed states. In order to obtain the final baryon asymmetry $\delta Y_B$ we evaluate $-\frac{2}{3} \sum_{l=e,\mu,\tau} \delta Y_l$ \cite{Dev:2014laa}.

The most general form for the diagonal matrix elements of $\mathcal{E}_N$  is
\begin{align}
	 \mathcal{E}_{N,\alpha\alpha} \equiv \frac{1}{n_{eq}^N} \int\limits_0^\infty \frac{\dd k\, k^2}{2\pi^2} \frac{\sqrt{k^2+m_{N_\alpha}^2}}{\exp\bigg[\sqrt{k^2+m_{N_\alpha}^2}\,z/T_s\bigg]+1},
\label{eq:E_N}
\end{align}
with  
\begin{align}
& n_{eq}^N = \frac{g}{2\pi^2}\int\limits_0^\infty \dd p \frac{p^2}{\mathrm{exp}\bigg[\sqrt{p^2+m^2}\,z/T_s\bigg]+1}, 
 & \rho_{eq}^N=\frac{n_{eq}^N}{s},
\label{eq:neq_full}
\end{align}
where $g=1$ for right-handed neutrinos. In the relativistic case, the expression in \cref{eq:E_N} reduces to \cite{Hambye:2017elz} 
\begin{align} 
  \mathcal{E}_N \approx  \frac{0.46 \, z}{2 T_{\text{s}}} m_{N_2}^2 \begin{pmatrix}
    1 &  0  \\
     0 &  (1+\delta_M)^2
  \end{pmatrix},
\end{align} 
where we introduced the level of degeneracy between the masses of $N_2$ and $N_3$ defined via relation $m_{N_3} = m_{N_2}(1+\delta_M)$. Note that, as we will show later explicitly, $\delta_M \lesssim 10^{-8}$ for a successful leptogenesis \cite{Hambye:2017elz}. 

In the numerical evaluation, we have employed the general expressions for all the terms involving right-handed neutrinos  given in \cref{eq:rate_eq,eq:rate_eq_bar,eq:rate_eq_asym}, i.e. we did not take the relativistic approximation. This is because
our findings indicate that successful leptogenesis occurs when $m_{N_{2,3}}\gtrsim T_s$, for which the relativistic approximation does not hold.
In contrast, the equilibrium number density for SM leptons is evaluated  in the relativistic approximation
\begin{align}
n_{eq}^L=\frac{3\, \zeta(3)}{2\pi^2}\left(\frac{T_s}{z}\right)^3,
\end{align}
since the heaviest SM lepton is two orders of magnitude lighter with respect to $T_s$. 
In \cref{eq:rate_eq,eq:rate_eq_bar,eq:rate_eq_asym} we denote the reaction densities for ARS and $\Sigma$ decays with $\gamma^{LC}$ and $\gamma^{LV}$, respectively, where $LC$ and $LV$ are abbreviations for the lepton number conserving and violating processes. 
Washout terms  are labeled with extra subscripts - $WC$ or $WQ$ where the former (latter) indicates lepton asymmetry loss due to classical (quantum) effects. The derivation of the reaction densities and the washout terms is outlined in \cref{app:phase_space_gamma}. Here, we would like to emphasize the scalar mass treatment in such calculation. First of all, in evaluating the integrals over the phase space,  for simplicity  and the possibility of analytical evaluation  we take $m_\pm=m_S=m_A$. We have nevertheless checked numerically that introducing the splitting between these masses, arising from $\lambda_{4,5}\neq 0$ (see \cref{eq:ms}), yields an insignificant change of reaction densities and washout terms which does not influence the final value of baryon asymmetry.

 In addition to the bare masses for new scalars, which are $\mathcal{O}(10^2-10^3)$ \,GeV, the thermal corrections are important. 
We therefore solve \cref{eq:rate_eq,eq:rate_eq_bar,eq:rate_eq_asym} in  two different temperature regimes. For the low temperatures, we neglect thermal corrections and use 
\begin{align}
m_\pm^2 = \mu_2^2 + \lambda_3 v^2,
\label{eq:bare}
\end{align}
 while for high temperatures we adopted  the expression for  thermal corrections, calculated in Refs. \cite{Gil:2012ya} and \cite{Blinov:2015vma}
\begin{align}
m_\pm^2 = \left( \frac{3\lambda_2 + 2\lambda_3 + \lambda_4}{12} + \frac{3g^2 + g'^2}{16}\right)\frac{T_s^2}{z^2},
\label{eq:thermal_scalar_mass}
\end{align}
where $g$ and $g'$ are $SU(2)_L$ and $U(1)_Y$ coupling constants, respectively. In total, we are using the following prescription for the scalar mass
\begin{align}
m_\pm^2 = \begin{cases}
 \left( \frac{3\lambda_2 + 2\lambda_3 + \lambda_4}{12} + \frac{3g^2 + g'^2}{16}\right)\frac{T_s^2}{z^2} + \mu_2^2 + \lambda_3 v^2, & \mathrm{if}\;\left( \frac{3\lambda_2 + 2\lambda_3 + \lambda_4}{12} + \frac{3g^2 + g'^2}{16}\right)\frac{T_s^2}{z^2} \geq 3\left( \mu_2^2 + \lambda_3 v^2 \right), \\
\,\, \mu_2^2 + \lambda_3 v^2, & \mathrm{otherwise} .
\end{cases}
\label{eq:thermal_mass_regimes}
\end{align}
The boundary temperature separating  the two regimes is determined by the condition that the thermal mass is equal to 3 times the bare mass. 
We checked numerically that varying the boundary temperature by $\mathcal{O}(1)$ numbers does not affect the final results.
 Below  this boundary we drop thermal corrections to the leptons, considering them to be effectively massless at low temperatures. Let us note that in both regimes, the phase space for $\Sigma$ decays is kinematically open, i.e. scalar masses are always larger than the total mass of decay products (right-handed neutrino and SM lepton). We would also like to stress that we have checked that in the relativistic limit our results match those presented in Ref. \cite{Hambye:2017elz}. 

\begin{figure}[ht!]
	\includegraphics[width=0.45\textwidth]{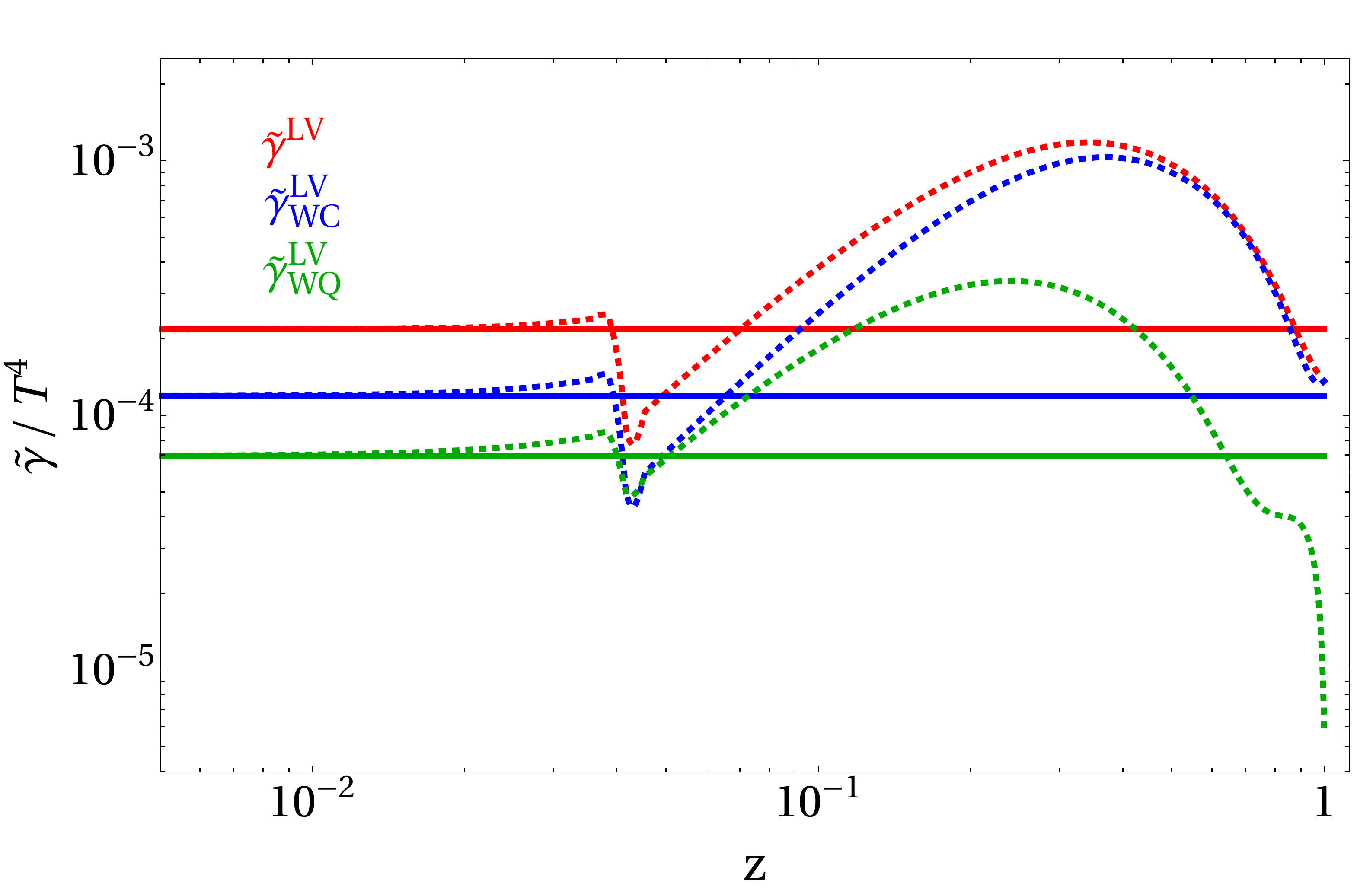}
	\includegraphics[width=0.45\textwidth]{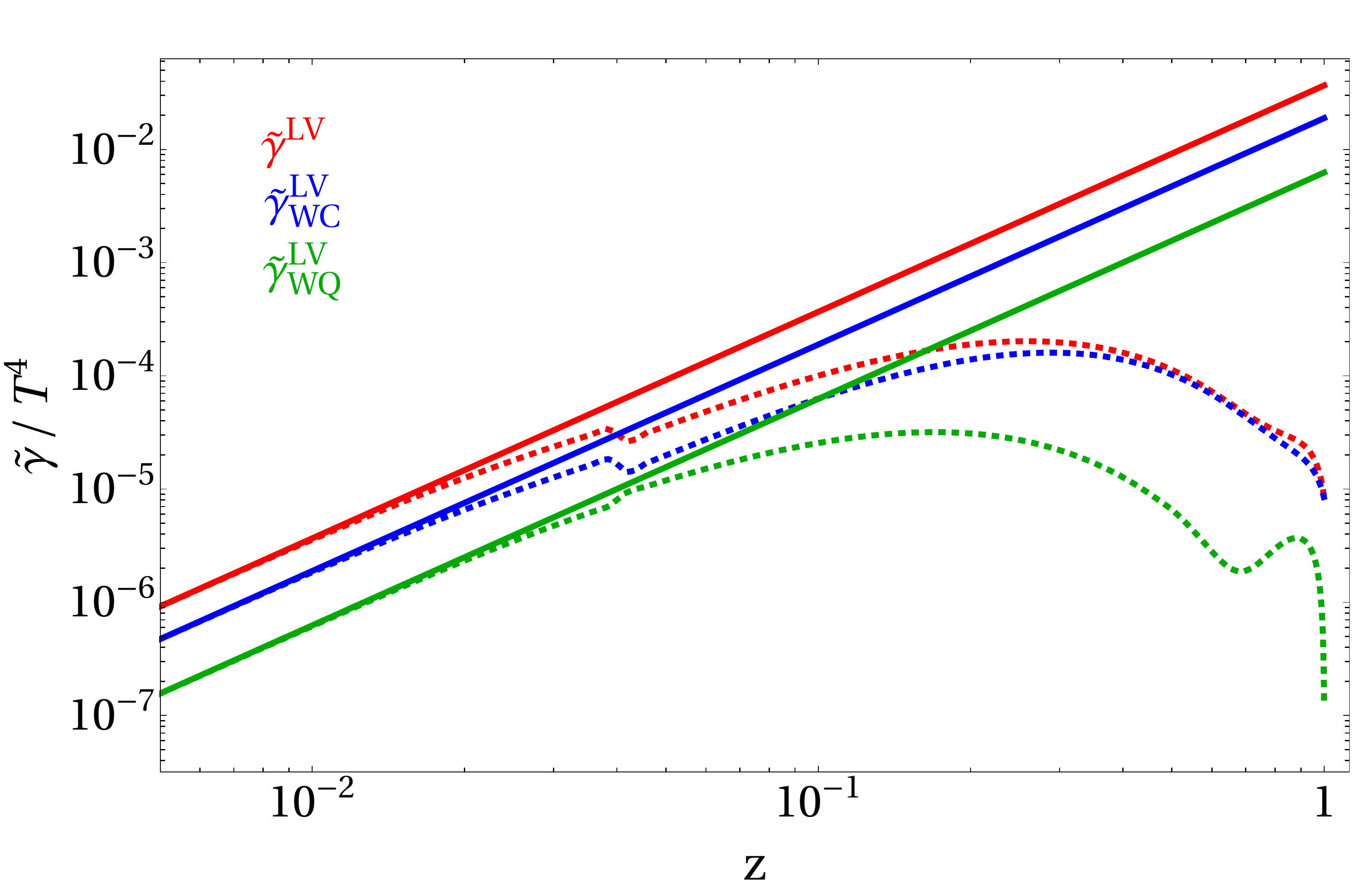}
	\caption{Temperature dependence of $\gamma^{LC}$ and $\gamma^{LV}$ (see \cref{app:phase_space_gamma}) with stripped-off Yukawa couplings (denoted with tilde). We take $\mu_2 = 1\TeV$, $m_{N_2}=400$ GeV and the values of the scalar couplings are given in \cref{tab:scalars}. The $T^4$ dependence is factored out in both panels. In both panels, the dashed (solid) lines correspond to the regime where bare (thermal) mass contributes. At $z\approx 0.05$ thermal corrections are as big as the bare mass, indicating a change of considered regime.}
	\label{fig:Fig_gamma}
\end{figure}

The temperature dependence of the $\gamma$ terms (reaction densities and washout terms) is shown in  \cref{fig:Fig_gamma} where we compare the relativistic regime (neglecting all bare masses and $m_N$, solid lines), with the general aforementioned approach (taking the scalar mass as given in \cref{eq:thermal_mass_regimes} and accounting for the effects from non-zero right-handed neutrino masses, dashed lines). We note the suppression of $\gamma^{LV}$ terms at low temperatures. This is because, in the absence of thermal effects, it is much less probable to have an on-shell mediator particle \cite{Hambye:2016sby}. The $\gamma^{LC}$ factors  feature an opposite effect -  get larger at $z\sim 0.1$ where the phase space for scattering processes between $L$ and $\Sigma$ increases. This is because $m_\pm$ is fixed to the bare mass and the lepton mass is gradually decreasing due to the dropping temperature. We also observe that all $\gamma$ terms in both panels become  Boltzmann suppressed at $z\sim 1$. 

\begin{table}[ht!]
\Large
	\begin{tabular}{c|c}
		quartic coupling\,\, & \,\,value \\
		\hline\hline
		$\lambda_2$ & $0.3$\\
		$\lambda_3$ & $-0.27$\\
		$\lambda_4$ & $2$\\
		$\lambda_5$ & $2$
	\end{tabular}
	\caption{Values of the scalar couplings motivated by a maximal reduction of the strength of $y_{2,3\,\alpha}$. Values are in accord with the limits coming from the  requirement for the stability of the Higgs potential (see \cref{subsec:limits}), as well as perturbativity.}
	\label{tab:scalars}
\end{table} 

 Radiative generation of neutrino masses introduces a suppression factor of $~\lambda_5^2/16\pi^2$ which lifts up the overall strength of ($y_{2\alpha}, y_{3\alpha}$) in comparison to the corresponding values in seesaw type-I model. Numerically, these couplings can not be made smaller than $\mathcal{O}(10^{-6})$ in our model.
 
Such interaction strength may pose a problem for generating observed BAU due to strong washout of the generated lepton asymmetry at late times. Let us note that, in principle, leptogenesis can be successfully achieved with such large couplings \cite{Hambye:2017elz} when $\xi$ (see \cref{eq:R}) is large. In addition, as pointed out in Ref. \cite{Shuve:2014zua}, the large imaginary entries (induced by $\xi$) may lead to certain cancellations between different terms in the Boltzmann equations, preventing the generated lepton asymmetry from strong washout effects. However, this reasoning can not be applied to our model because pushing $\xi$ to large values drastically increases the strength of washout effects. In fact, even the  asymmetry produced at very late times will be washed out efficiently before sphaleron decoupling.
 
In order to reduce the Yukawa couplings as much as possible we choose the quartic couplings as given in \cref{tab:scalars}, consistent with the limits from the stability of the Higgs potential (see \cref{subsec:limits}).  With couplings chosen in such a way we find that successful leptogenesis may only be generated with $m_{N_2}\gtrsim \mathcal{O}(10^2)$ GeV. For the right-handed neutrino masses smaller than $T_s$, the washout effects at $z\gtrsim 0.1$ remove the vast majority of the generated asymmetry. In contrast, if $m_{N_2}\gtrsim T_s$, the right-handed neutrino abundance becomes Boltzmann suppressed at $z\gtrsim10^{-1}$ which strongly suppresses washout integrals given in \cref{app:phase_space_gamma}. While at such late times the asymmetry production is suppressed in a similar way,  we find that the strong production at $z\ll 1$ suffices for generating the observed $\delta Y_B=0.86\times 10^{-10}$ \cite{Ade:2015xua} baryon asymmetry. Throughout the paper, without the loss of generality, we use only positive values of $\xi$. This parameter enters exponentially ($\exp^{\pm \xi}$) in the Yukawa matrix and thus its negative values would not change qualitatively the overall picture for BAU generation.

\begin{figure}[ht!]
	\includegraphics[width=0.49\textwidth]{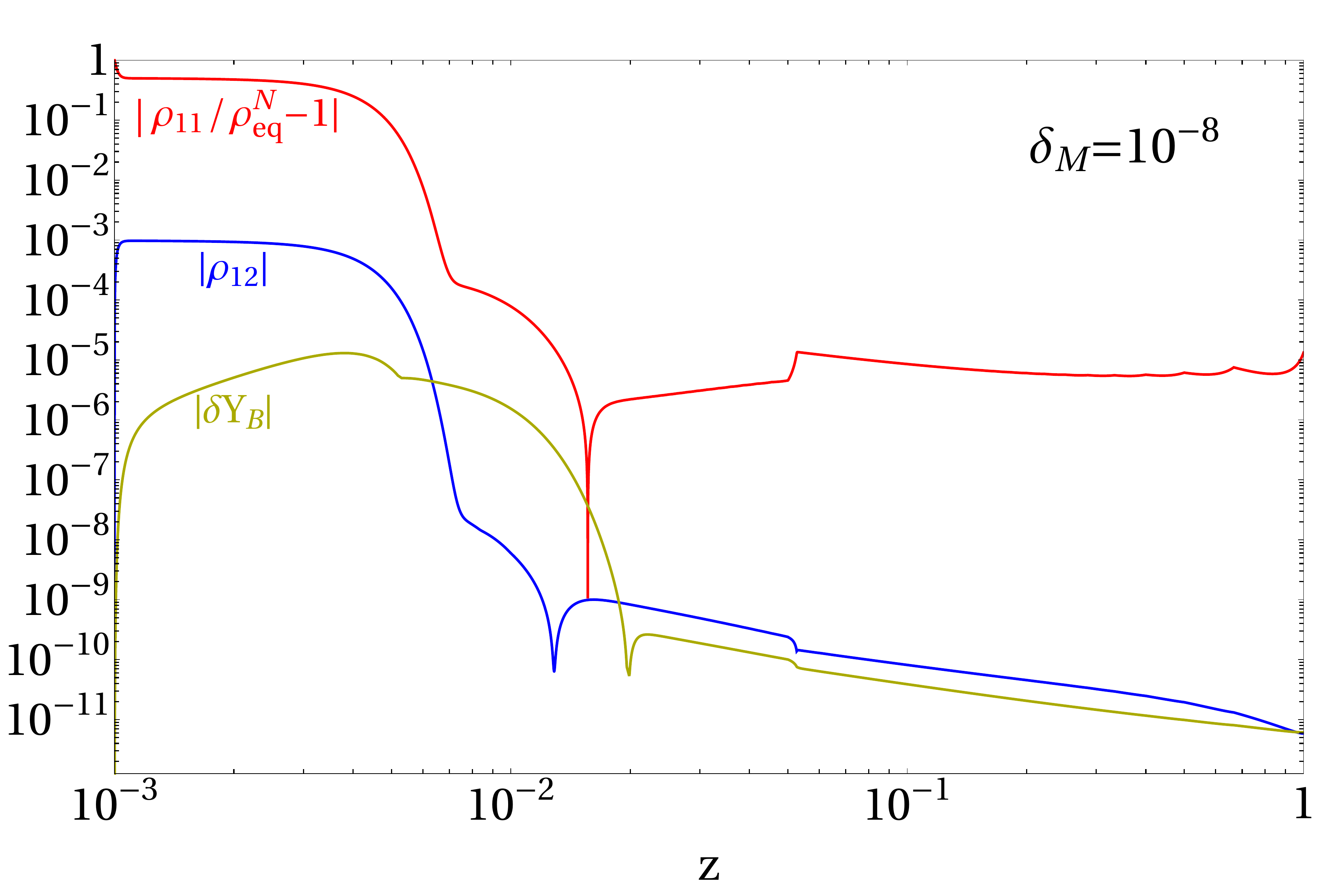}
	\includegraphics[width=0.49\textwidth]{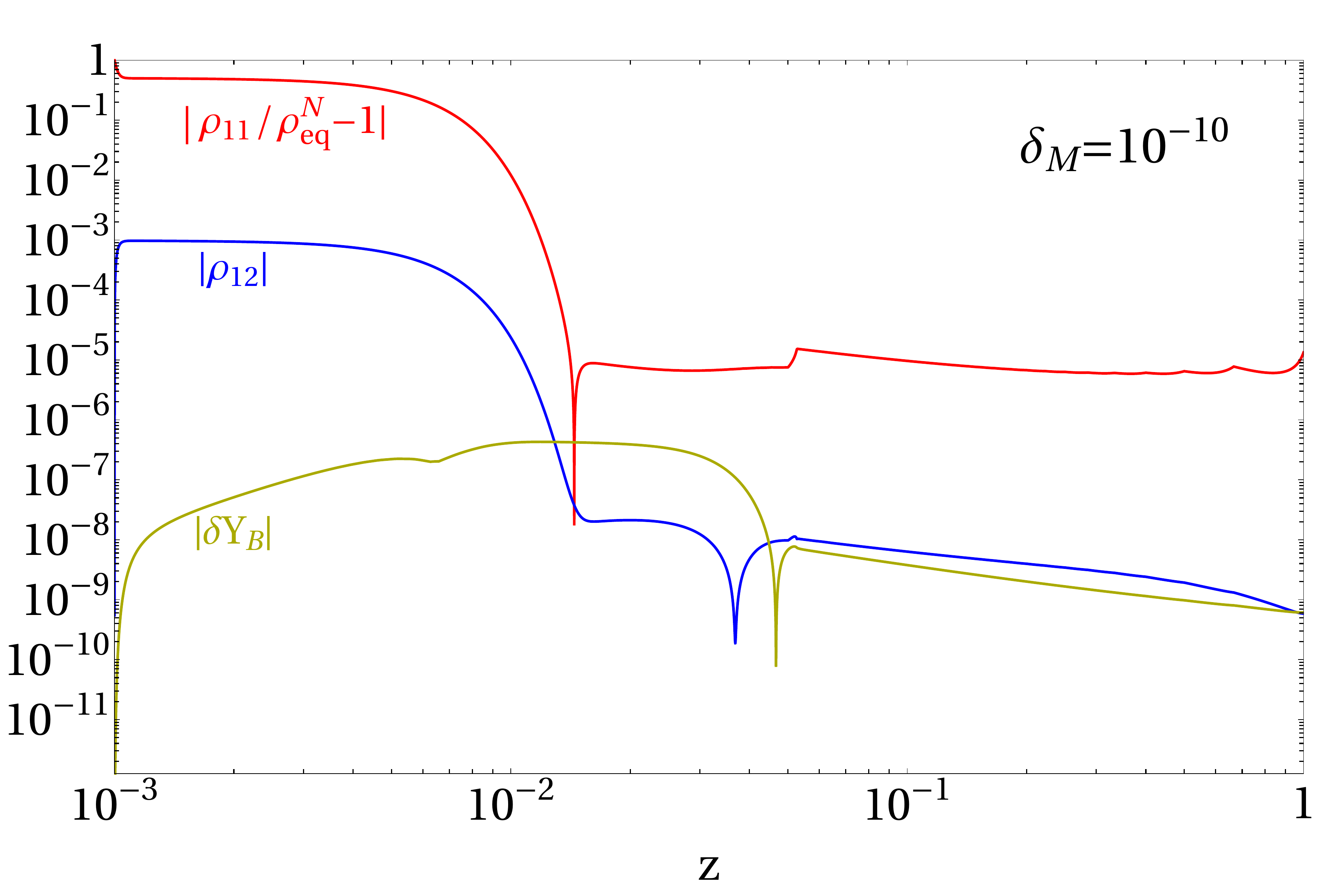}
	\caption{Evolution of $\rho_{11}(z)$, $\rho_{12}(z)$ and $\delta Y_B(z)$ for two different values of $\delta_M$. We have fixed $\omega=\pi/4$, $\delta= - \pi/2$, $\xi = 2$ and $\alpha_1=\alpha_2=0$. The asymmetry for $\delta_M=10^{-8}$ is produced earlier but also more strongly washed out with respect to $\delta_M=10^{-10}$ case. We found no significant dependence of the generated asymmetry on the values of Dirac ($\delta$) and Majorana ($\alpha_1$ and $\alpha_2$) CP phases.}
	\label{fig:results_asymm}
\end{figure}

 In \cref{fig:results_asymm} we show the deviation from the equilibrium value of $N_2$ number density ($\rho_{11}(z)/\rho_\text{eq}^N-1$) and the mixing between $N_2$ and $N_3$ ($\rho_{12}(z)$). The evolution of the baryon asymmetry $\delta Y_B$ is also presented for $m_{N_2}=200\GeV$, $\mu_2=800\GeV$ and with the following values of angles and phases $\omega=\pi/4,\,\delta=-\pi/2,\,\alpha_1=\alpha_2=0$. The kink at $z\approx 0.05$ is due to the change of regimes for the thermal masses (see \cref{eq:thermal_mass_regimes}). The narrow feature at slightly smaller temperatures indicates the sign change of $\delta Y_B$. A significant deviation  of  $\rho_{11}$ and $\rho_{22}$ from $\rho_\text{eq}^N$ as well as avoiding very tiny off-diagonal matrix elements ($\rho_{12}$ and $\rho_{21}$) are crucial for the successful leptogenesis. This is only ensured for  $m_{N_2}\gtrsim T_s$, as discussed above. We would like to stress that there is no significant dependence of the generated lepton asymmetry on $\alpha_1$ and $\alpha_2$. In addition, note that $\delta_M$ has to be  tiny for generating the asymmetry, implying strong level of degeneracy between heavier right-handed neutrino masses.
 From the shape of $\rho_{12}(z)$ and $\delta Y_B(z)$ curves for 
$\delta_M=10^{-8}$ (left panel) and $\delta_M=10^{-10}$ (right panel) we infer 
that in the former case, the washout effects are effective throughout a longer time period. This can be understood from the temperature scale $z_\mathrm{osc}$ at which the oscillation among right-handed neutrinos is most effective. It is given by \cite{Akhmedov:1998qx}
\begin{align}
z_\mathrm{osc}=T_s \,\left(2\sqrt{45/(4\pi^3 g_*)}\,m_{N_2}\,\delta_M\,M_{\mathrm{Pl}}\right)^{-1/3}.
\end{align}
For smaller $\delta_M$, the asymmetry is produced later which leads to a weaker washout and consequently larger $\delta Y_B$. 

\begin{figure*}[t!]
	\centering
	\subfigure[ \,$\delta Y_B$ in ($\xi,m_{N_2}$) plane for $\lambda_5= 2$.]{\label{fig:6a}\includegraphics[width=60mm]{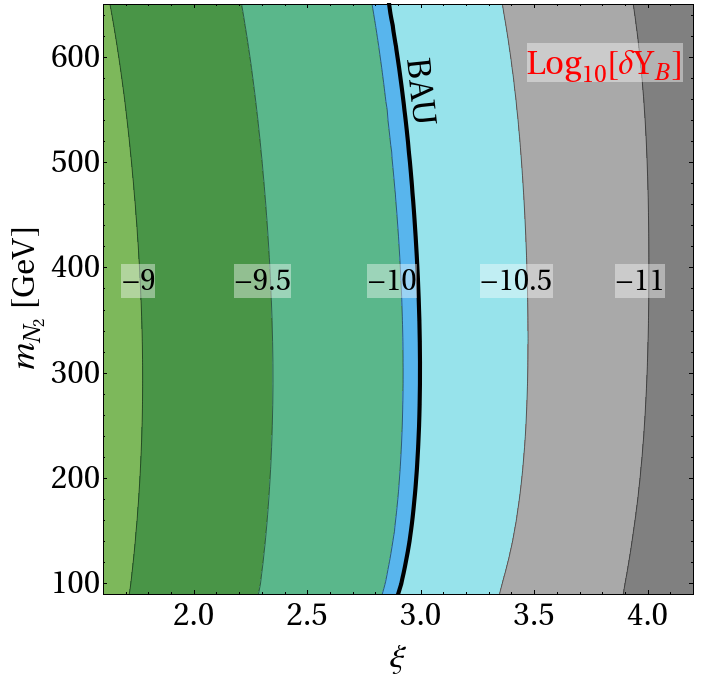}}
	\subfigure[\,$\delta Y_B$  in ($\xi,m_{N_2}$) plane for $\lambda_5 = 0.01$.]{\label{fig:6b}\includegraphics[width=60mm]{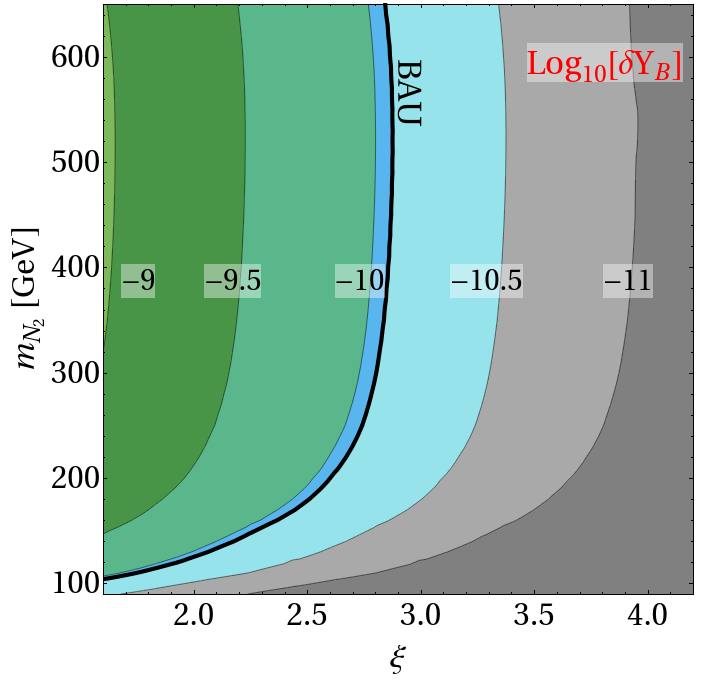}}
	\subfigure[\,$\delta Y_B$  in ($\log \lambda_5,m_{N_2}$) plane, $\xi=1$.]{\label{fig:6c}\includegraphics[width=60mm]{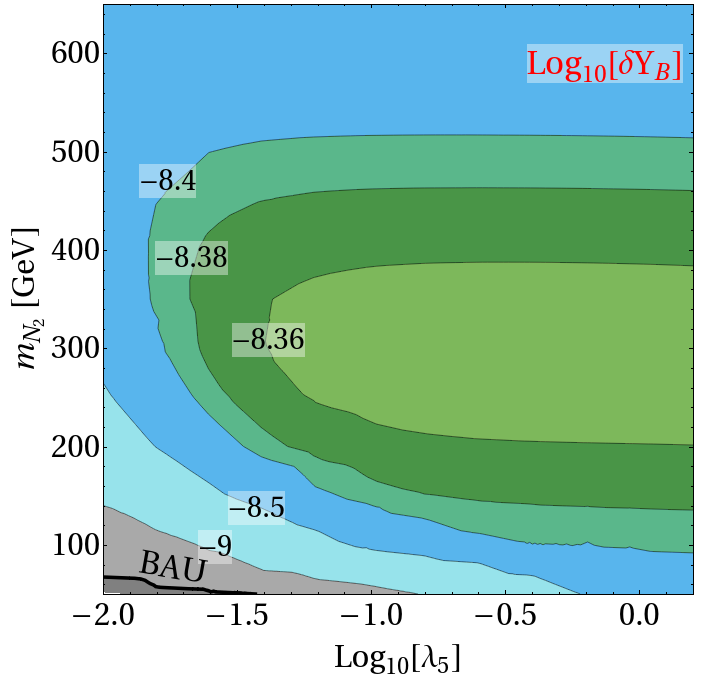}}
	\subfigure[\,$\delta Y_B$  in ($\xi,\log \lambda_5$) plane, $m_{N_2}= 200\GeV$.]{\label{fig:6d}\includegraphics[width=60mm]{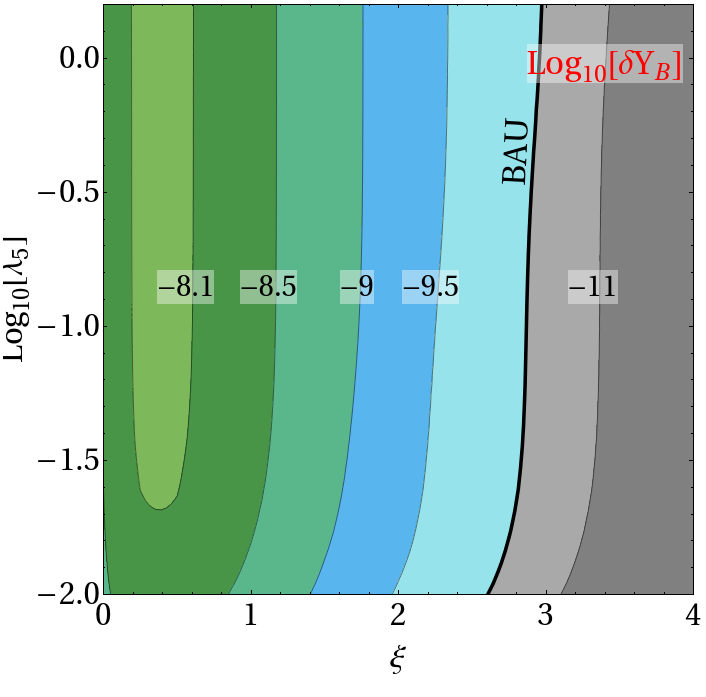}}
	\caption{$\delta Y_B$ shown for various choices of parameters. The level of degeneracy between $N_2$ and $N_3$ masses is fixed to $\delta_M=10^{-10}$ in all panels. The solid black line indicates the value of the observed baryon asymmetry of the Universe.}
	\label{fig:Parameter_scans}	
\end{figure*}

In \cref{fig:6a,fig:6b} we show the generated baryon asymmetry in $\xi-M_{N_2}$ plane for two fixed values of $\lambda_5$. From this figure one can also qualitatively infer the dependence of the generated baryon asymmetry on the Yukawa interaction strength since different values of $\lambda_5$ imply different $y_{2,3\,\alpha}$.
As expected, there is only a mild dependence on $m_{N_2}$ which  does not alter the Yukawa strength dramatically.   
 From  \cref{fig:6c} one can deduce that there is a saturation effect for $\delta Y_B \gtrsim 4\times 10^{-9}$, where the dependence of the asymmetry on $\lambda_5$ flattens. Furthermore, one should note that in case of large $\lambda_5$, $\delta Y_B$ is generally very weakly dependent on this quartic coupling. Finally, \cref{fig:6d} summarizes the interplay between $\lambda_5$ and $\xi$. Intuitively, it is clear that the largest asymmetry production happens for low $\xi$ and high $\lambda_5$ values, whereas going to the opposite regime favors  washout, hence leading to a reduced final asymmetry. Interestingly, $\xi$ has a stronger impact than the quartic coupling, despite the fact that in the considered parameter range both quantities vary the size of the Yukawa coupling by almost two orders of magnitude. The reason for this behavior is that while $\lambda_5$ just sets the overall strength of the Yukawa couplings, $\xi$ enters in the Yukawa matrix in a specific pattern, leading to rather non trivial effects. 
 
\begin{figure}[ht!]
	\centering
	\includegraphics[width=0.55\textwidth]{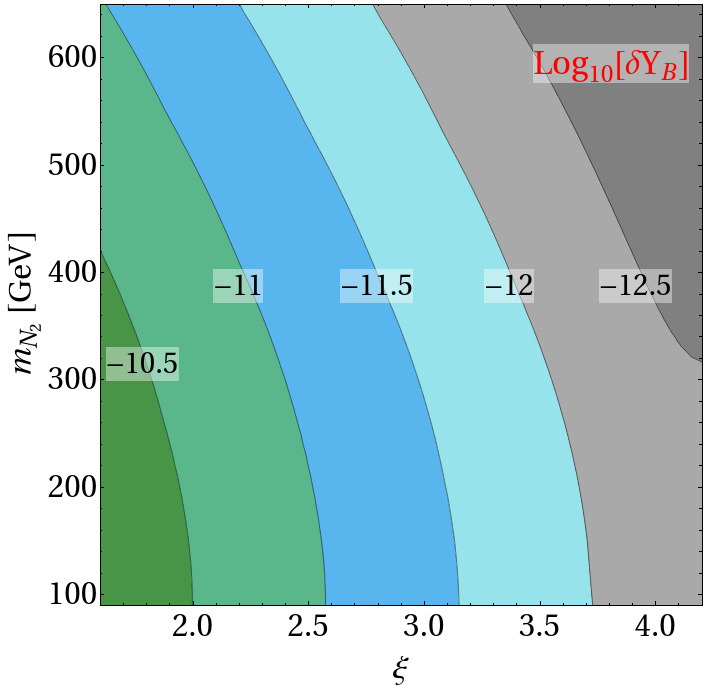}
	\caption{Same as \cref{fig:6a} but with only ARS contribution taken into account. This generically leads to a smaller final asymmetry production as can also be seen in \cref{fig:Fig7b}.}
	\label{fig:Fig7a}
\end{figure}

 From \cref{fig:Fig7a} we observe that when considering only ARS, $|\delta Y_B|$ decreases roughly by one order of magnitude with respect to the values from \cref{fig:6a}. What can also be inferred is the stronger dependence on $m_{N_2}$ in absence of scalar decays.
 
\begin{figure}[ht!]
	\centering
	\includegraphics[width=0.75\textwidth]{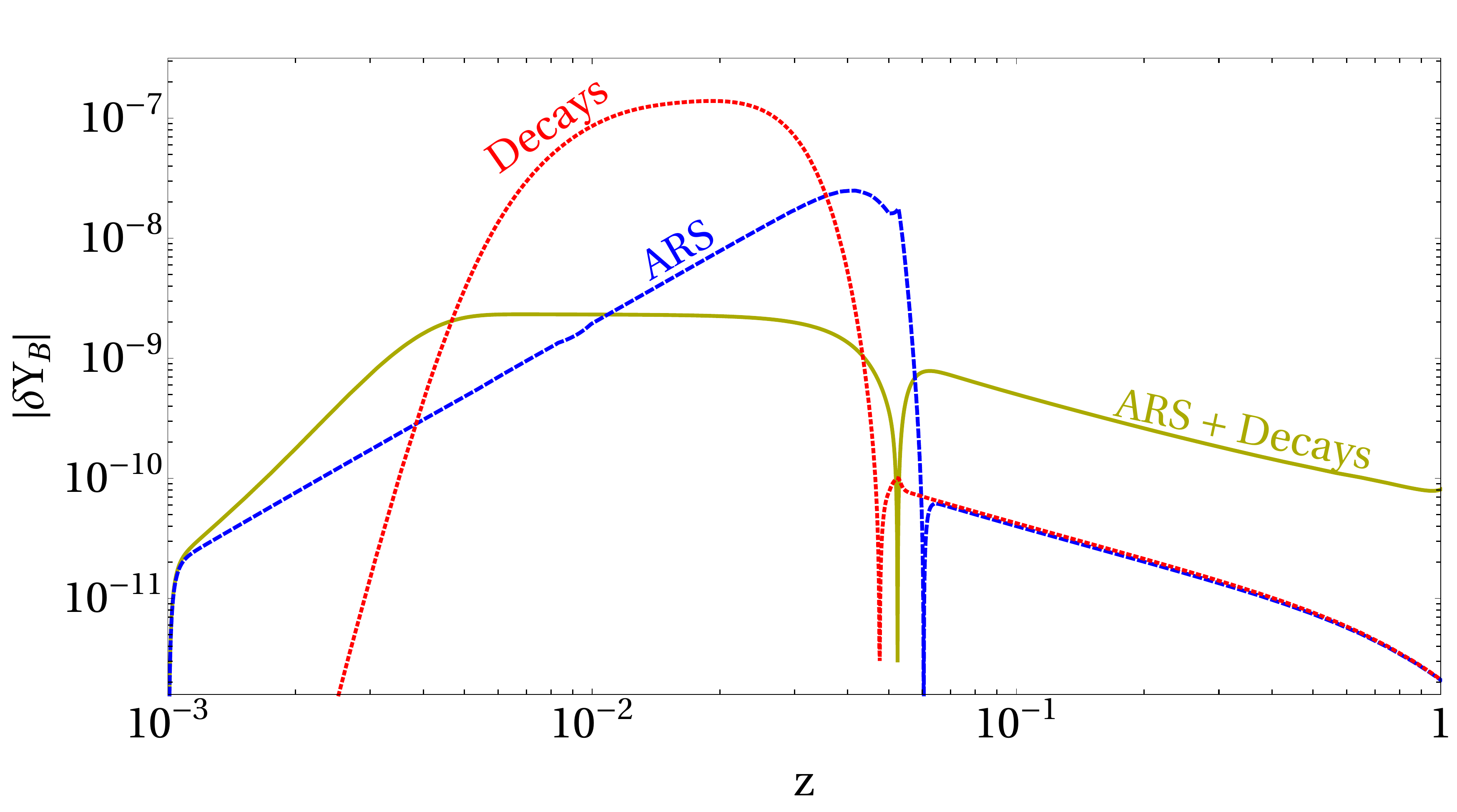}
	\caption{Time evolution of $|\delta Y_B|$ for the cases of $\Sigma$ decays (red dotted) and ARS (blue dashed) shown together with the general case when the both production mechanisms are included (yellow solid). We take $\xi = 3$ and $m_{N_2}=400\GeV$. It is evident that the combination of both production mechanisms suffices to account for BAU, in contrast with the individual contributions.}
	\label{fig:Fig7b}
\end{figure}

We find that it is very important to  account for the asymmetry production via new scalar decays in addition to ARS. Actually, turning either of the two production mechanisms off yields a loss of a substantial amount of the generated baryon asymmetry,  as evident from \cref{fig:Fig7b}.  Obviously, there is a strong interplay between both regimes, because their individual contributions are rather small and do not trivially add up to yield the final result. Each of the regimes individually leads to a similar final asymmetry contribution. Initially, the asymmetry generation is governed exclusively by the ARS mechanism and $\Sigma$ decays start to compete at $z\approx 0.005$. Interestingly, the intermediate asymmetry produced by $\Sigma$ decays is one order of magnitude larger compared to the one in ARS, but tends to get washed out more strongly. At $z\approx 0.05$, $\delta Y_B$ changes its sign and later fades away due to washout effects. At even smaller temperatures all regimes feature similar asymmetry evolution (governed by washout effects) until sphaleron processes eventually decouple.
Again, including both regimes does not translate into summing up their individual contributions. The asymmetry generation in the presence of both mechanisms shows a qualitatively different behavior, featuring a peak at larger temperatures. Non-linear combination of ARS and scalar decays explicitly  demonstrates the absence of the weak-washout regime in which the contributions would add trivially.

\section{Identifying the viable joint parameter space for DM and leptogenesis}
\label{sec:combined}

In \cref{sec:production,sec:leptogenesis} we separately scrutinized DM and BAU production. Therefore, the question arises whether it is possible to find regions in the parameter space for which the observed values of DM relic abundance and $|\delta Y_B|$ are simultaneously reached. Generally, these two mechanisms have conflicting requirements on the strength of the Yukawa coupling. In order not to overproduce DM from decays of $N_2$ (\cref{sub:N2_decay}), sufficiently large Yukawa interactions are required. On the other hand, leptogenesis relies on weak interactions as otherwise the washout effects would easily destroy any generated lepton asymmetry.

A key method for curing this contrast is to impose coannihilations between the right-handed neutrinos and the scalars (especially the lightest scalar $\sigma^\pm$) by choosing them to have similar masses.  The coannihilation case was already discussed in \cref{sec:production} (see in particular 
\cref{fig:Freeze_Out_N2}).
Such a regime opens up new scalar annihilation channels  which do not rely on the strength of the second and third generation Yukawa couplings. Therefore, a huge suppression of the relic density can be achieved  for Yukawa couplings  set low enough to generate a significant lepton asymmetry.

We now revisit BBN constraints on the $N_2$ decays introduced in \cref{subsec:limits}. By requiring $N_2$ to decay before $t_\text{BBN}\sim 1$ sec we obtain 
\begin{align}
|y_{2\alpha}|^2 \gtrsim 6.3\cdot10^{-7}\left( \frac{m_\pm}{1\TeV} \right)^4 \left( \frac{1\TeV}{m_{N_2}} \right)^5 \left( \frac{10^{-8}}{|y_1|} \right)^2.
\label{eq:BBN_bound}
\end{align}   
After a more careful analysis of the processes influencing the primordial abundances of light elements this limit gets significantly relaxed.
Following Ref. \cite{Kawasaki:2017bqm}, we infer that for $Y m_{N_2}\lesssim \mathcal{O}(10^{-9})$, where $Y$ is the DM yield (see \cref{eq:yield_boltzmann}),  the BBN limit given in \cref{eq:BBN_bound} is relaxed by roughly three orders of magnitude. This is because, in our model, $N_2$ has only leptonic decays and the strongest effect from charged leptons on the primordial abundances of the light nuclei is coming from the photodissociation process which is most effective at approximately $1000$ seconds after the Big Bang.

\begin{figure}[ht!]
	\centering
	\includegraphics[width=0.48\textwidth]{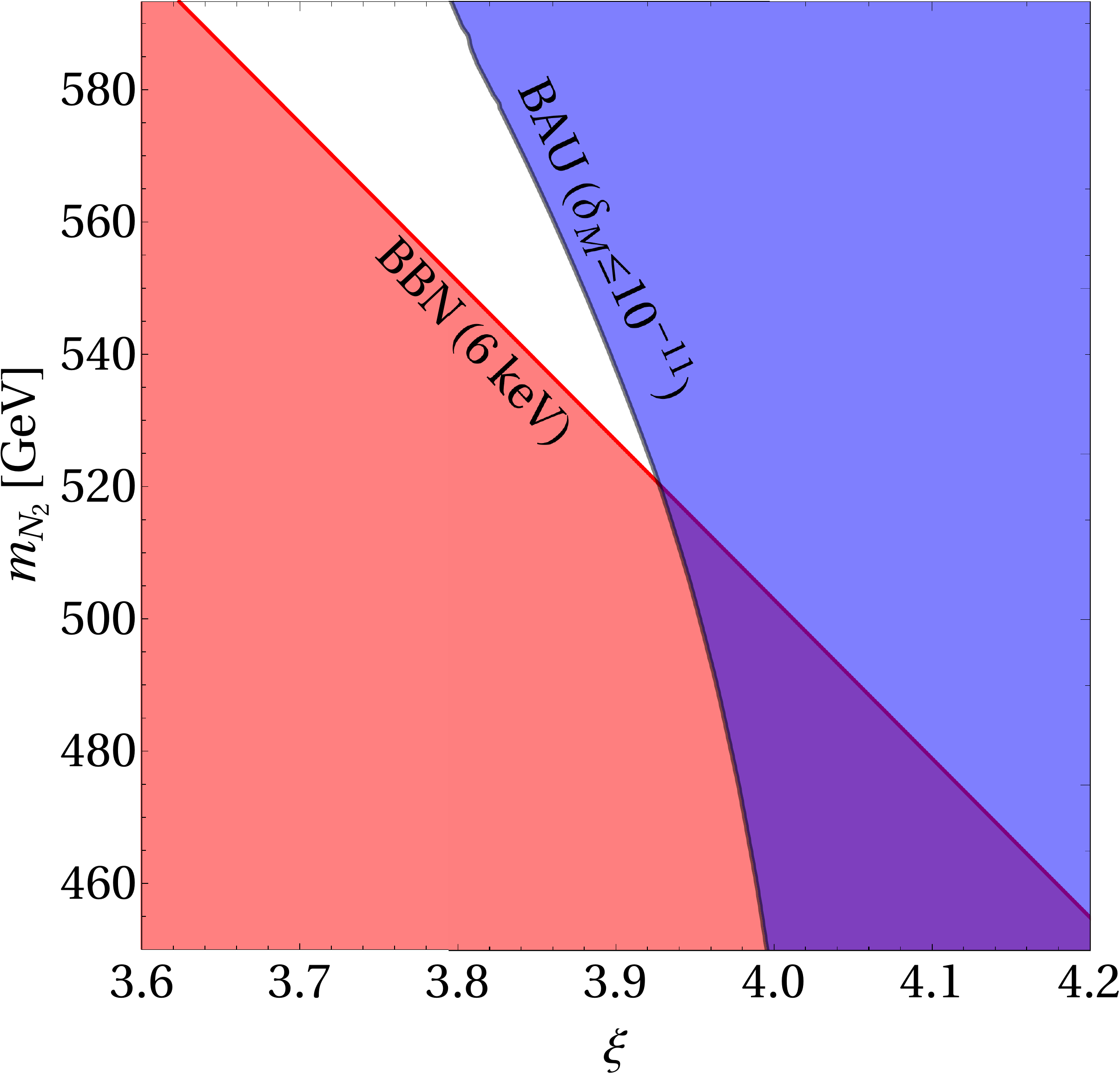}
	\includegraphics[width=0.48\textwidth]{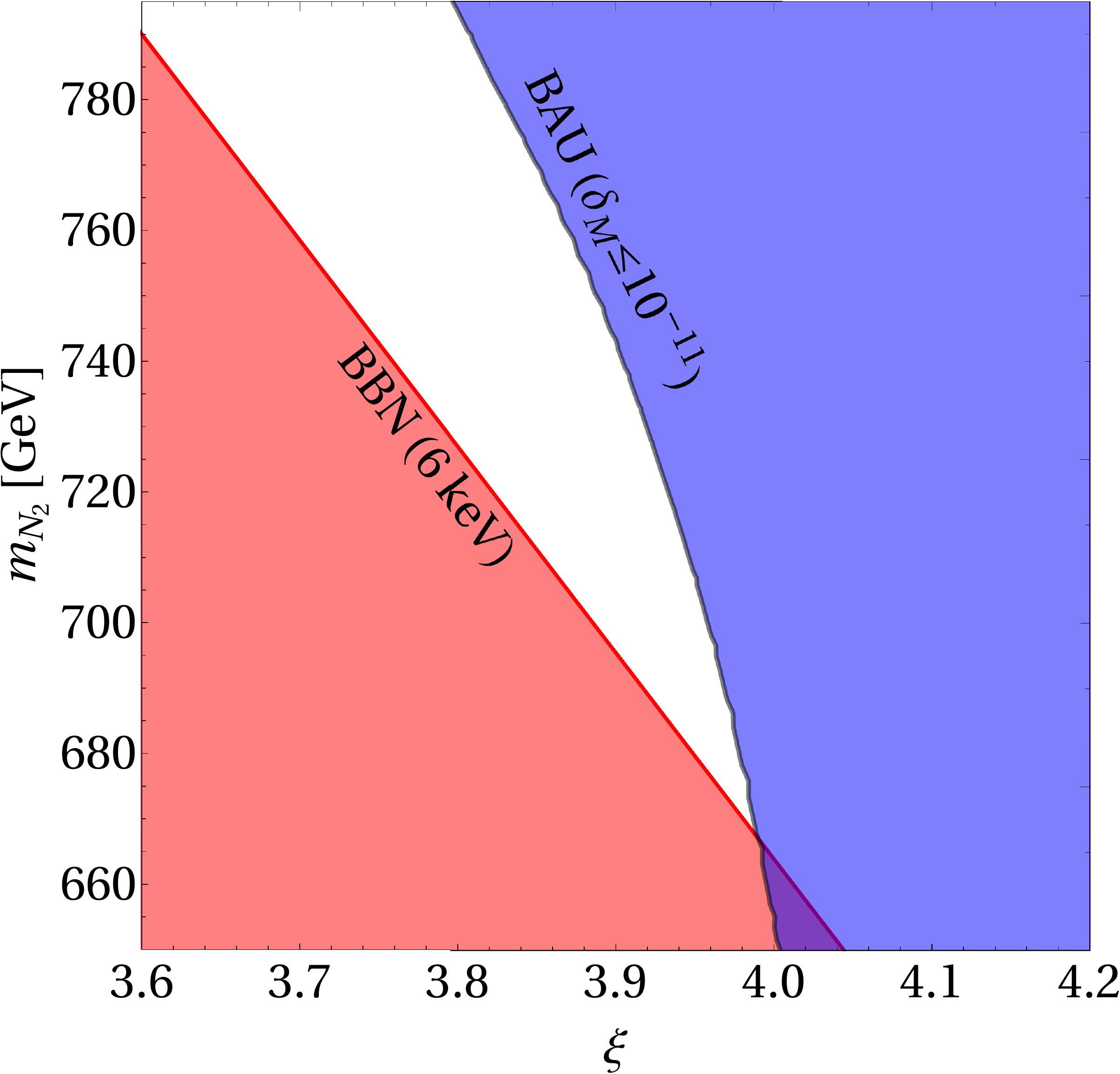}
	\caption{
	Allowed region in the ($\xi,m_{N_2}$) parameter space for fixed $\mu_2=600\GeV$ (left) and $\mu_2=800\GeV$ (right). The BBN exclusion limits are shown in red, while the blue shaded region does not produce a sufficiently large baryon asymmetry. We observe that there is a substantial region consistent with BBN limits in which the correct amounts of DM and baryon asymmetry can be obtained. }
	\label{fig:Fig8}
\end{figure}

We calculated the relic density $\Omega h^2_{N_2}$ for different choices of $\mu_2$ and
also evaluated $\delta Y_B$ in a range of $m_{N_2}$ and $\xi$, taking into account $m_{N_2} < m_\pm$. Then, we determined which region in the parameter space is ruled out due to low $y_{2\alpha}$. We conservatively adopted the limit for the most stringent BBN constraint on pure leptonic decays  ($\tau^+\tau^-$, see Ref. \cite{Kawasaki:2017bqm}). In \cref{fig:Fig8} we present the results for two different choices of $\mu_2$. In both panels, the blue region is indicating excluded parameter space due to the insufficient amount of generated asymmetry and the red region is excluded by BBN.\\

There are several things we would like to point out. First, the final baryon asymmetry only mildly depends on the involved particle masses and in general larger mass scales will only lead to a  slight decrease of $\delta Y_B$ as can be seen from the slope of the BAU line in \cref{fig:Fig8}. In contrast, choosing a higher mass scale weakens the BBN bounds, e.g. for a DM mass of $6$ keV we need  $\Omega h^2_{N_2}\lesssim 8.5$ for $\mu_2=600\GeV$, while for $\mu_2=800\GeV$ the limit is relaxed  to $\Omega h^2_{N_2}\lesssim 88.7$. 
These values also suggest that the $N_2$ decay contribution  ($\Omega h^2_{N_2\to N_1}$) to the DM relic abundance is negligible and thus $y_1$ is fixed by the requirement that DM is completely produced via $N_1$ freeze-in. Larger scalar masses  lead to stronger Yukawa interactions in order to maintain the correct relic density (see \cref{{eq:relic_abundance2}}).

In \cref{fig:Fig8}, the blue region indicates the region excluded for $\delta_M \leq 10^{-11}$. A smaller level of degeneracy between right-handed neutrinos, such as $\delta_M=10^{-10}$,  would correspond to the shift  of this exclusion to the left (toward BBN bound). For $\delta_M\geq 10^{-10}$ and $m_{N_1}\sim 6\keV$ we find no parameter space free from either BBN or BAU exclusions. 

We have also observed the dependence of BBN limits on the maximum allowed DM mass. For $\mu_2$ in the range $\left[300, 1000\right]$ GeV, we found a maximal value of the DM mass of 9.4~keV, which is consistent with structure formation bounds. Generally, choosing higher degeneracies will open up the available parameter space, thus allowing larger DM masses. However, even in such cases we estimated the maximal allowed DM mass to be at most $\mathcal{O}(10)$ keV.   

In summary, we identified the parameter space in which the produced DM abundance and BAU are in accord with the observed values. We have seen that the most stringent constraints arise from BBN considerations, which can be relaxed by employing coannihilation processes between $N_{2,3}$ and $\Sigma$ which effectively put an upper bound on the allowed mass for the DM. We found that the DM production in our model is mainly driven by the freeze-in.

In this section we have shown that the three biggest challenges of particle physics, pointed out in  \cref{sec:intro}, can be successfully and simultaneously solved  within the considered model.

\section{ Detection Prospects}
\label{sec:detection}

In \cref{subsec:limits} we discussed the limits from structure formation on keV-scale DM. Here, we wish to point out that these limits are relaxed in this model which opens up some parameter space for such DM candidate. Let us illustrate how DM can be made colder, i.e. less constrained from structure formation considerations.
We discuss the most relevant case where the observable DM relic density
is  dominantly set through the freeze-in from the decays of heavy scalar particles (see \cref{sub:decay}), as the strong production from $N_2$ decays is in tension with BBN limits. 
The production of  DM occurs at the mass scale of decaying charged and neutral scalars. Due to the production at such high temperatures, cooling of  DM particles is efficient. Namely, the effective temperature of the DM sector, when compared to photons, is reduced by the amount of entropy dilution factor which is  $\approx 2.9$ \cite{Brdar:2017wgy,Heeck:2017xbu}.

Having the absence of X-ray signal, we reach the conclusion  that the testability of our model is currently limited  only to the searches at the LHC as well as the facilities probing lepton flavor violation processes. The limits coming from the latter  are  discussed
in \cref{subsec:limits} and consistently taken into account throughout the paper. In this section, therefore, we mainly comment on the LHC prospects,
which were studied for this model in \cite{Hessler:2016kwm} where
the dominant production of DM is assumed to be via freeze-in through scalar decays.  Among others, the authors are 
considering the regime where scalar particles are heavier than all three generations of right-handed neutrinos, which matches our setup.

The answer to the question which LHC search  has the strongest sensitivity depends on the mass difference between $m_{N_2}$ and $m_{N_3}$. In \cref{sec:leptogenesis}, we showed that baryogenesis can be achieved via generation of a lepton asymmetry, where the crucial ingredient was the approximate degeneracy between  $N_2$ and $N_3$ states. In that case, 
the LHC searches are limited to $\sigma^\pm \to N_{2,3} \, l_{\alpha}^\pm$, i.e. channels including charged leptons in the final state. Charged scalars $\sigma^\pm$ are produced \emph{in pairs} either via gluon fusion or Drell-Yan processes \cite{Hessler:2016kwm}, and therefore the expected signature consists of two prompt charged leptons and missing energy due to elusive right-handed neutrinos which 
are decaying to $N_1$ only after leaving the detector. 

Let us note that with an assumption of a complementary baryogenesis mechanism that goes beyond our framework (for instance decays of very heavy singlets \cite{Fukugita:1986hr} or electroweak baryogenesis \cite{Morrissey:2012db}) for which tiny $\delta_M$ is not required, the discovery  potential at LHC increases. Namely, if we assume  a mass gap between $N_2$ and $N_3$ to be $\gtrsim 10 \,\text{GeV}$ there is another viable search in this model -- displaced vertices \cite{Buchmueller:2017uqu}. Even though decays of heavier right-handed neutrinos into DM ($N_1$) do not happen within the detector, the decay $N_3\to N_2 l_\alpha^\pm l_\beta^\mp$ can be rapid enough. Then, the displaced leptons from the initial scalar decay and the subsequent $N_3$ decay may be observed. Limits for both leptons+$\slashed{E}_T$ and displaced vertices are presented in \cite{Hessler:2016kwm}. Let us emphasize that displaced vertex search 
can constrain $y_{3\alpha}$ and $y_{2\alpha}$ couplings up to the level of $10^{-4}$, which is two orders of magnitude stronger with respect to the limits from LFV assuming no strong hierarchy between the entries of the Yukawa matrix. The existence of the upper limit is a consequence of the requirement for the resolution of displaced vertices. It is worth mentioning that the limits from LHC searches attenuate very quickly above EW scale. For more details we refer the reader to Ref. \cite{Hessler:2016kwm}.

Let us also remark that the presence of extra charged scalars implies the radiative contribution to the decays of SM Higgs particle into a pair of photons or a photon and a $Z$ boson \cite{Swiezewska:2012eh,Arhrib:2012ia}. Specifically, for the diphoton channel, in order to get constructive (destructive) interference between SM and new physics contribution, one requires negative (positive) Higgs portal couplings. Significant deviation from measured values 
\cite{Patrignani:2016xqp} are achieved for rather large values of such couplings \cite{Carena:2012xa,Picek:2012ei} which are avoided in our analysis.

We finalize this section with a comment on the detection prospects for  $N_{2,3}$  discovery at future lepton colliders as well as SHIP \cite{Lantwin:2017xtc}, that were proposed recently \cite{Blondel:2014bra,Hernandez:2016kel,Antusch:2017pkq}. At lepton colliders, right-handed neutrinos are assumed to be generated in the process $e^+ e^- \to N_{2,3} \,\nu_\alpha$ \cite{Banerjee:2015gca}, for which the  mixing between active and right-handed states is required. 
The mixing is necessary also for SHIP where right-handed neutrinos would get produced in the decays of heavy hadrons. Hence, due to the absence of the mixing induced by an exact $\mathbb{Z}_2$ parity symmetry, right-handed neutrinos in this model are not testable at these facilities.

\section{Summary and Conclusions}
\label{sec:summary}

In this work we studied the generation of neutrino masses, dark matter and baryon asymmetry of the Universe within the so called 
Scotogenic model, where three right-handed neutrinos and an additional scalar doublet, all odd under a $\mathbb{Z}_2$ parity symmetry, are added to the SM.  Active neutrino masses are obtained radiatively via loops involving new particles. We considered a mass spectrum where all scalar masses are at or below the TeV-scale. Furthermore, we invoked a hierarchy in the right-handed neutrino mass spectrum, choosing $m_{N_1}\approx\mathcal{O}(1)$ keV and $m_{N_{2,3}}\approx \mathcal{O}(100)$ GeV, where the lightest state, $N_1$, is a keV-scale DM particle.

For DM production, we looked at two complementary contributions : First, we examined freeze-in of $N_1$ from the decays of new scalars. Second, we also took three-body decays of ``frozen-out" $N_2$ into account.  We were able to derive the correct DM density in a wide parameter region.

As the baryon asymmetry is concerned, we studied leptogenesis from the combination of right-handed neutrino oscillations and scalar decays.
 We showed that it is possible to derive a significant asymmetry in case of highly degenerated right-handed neutrino masses. We have established that it is crucial to set the mass of heavier two states to $\gtrsim \mathcal{O}(10^2)$ GeV  in order to prevent late time washout. 
 
 Finally, in finding the joint  parameter space for DM and BAU, we have shown that the BBN bound plays a major role and rules out a large portion of the parameter space, effectively forbidding DM mass to exceed $\sim 10$ keV. Nevertheless, by imposing coannihillations between fermions and scalars,  we were successful in  identifying the regions where both DM and BAU are produced in right amounts. Hence, in the considered model, we have simultaneously solved the three greatest mysteries in particle physics.
  
\section*{Acknowledgments}
We thank Joachim Kopp for the collaboration in the initial stage of this project.
We are greatly indebted to Daniele Teresi for cross checking our leptogenesis implementation in the initial phase of this work as well as  helping us understand several key leptogenesis features, in particular the production from scalar decays. VB would also like to thank
Mikhail Shaposhnikov and Kai Schmitz for very insightful and useful discussions on the presented model. Work in Mainz is supported by the Cluster of Excellence Precision Physics, Fundamental Interactions and Structure of Matter (PRISMA -- EXC 1098).
\appendix
\section{Phase space integration and reaction densities}
\label{app:phase_space_gamma}

Here we present the results for the reaction densities and washout terms, obtained by  performing the phase space integration. In the following, we make the abbreviation $E_i\equiv E_i(\textbf{p}_i)$ for 
SM lepton ($E_L$), right-handed neutrino ($E_N$) and scalar ($E_\Sigma$) energies. The right-handed neutrino and SM lepton momenta are denoted with 
$k$ and $p$, respectively.
We also use the approximation $m_{N_2}\simeq m_{N_3} \equiv m_{N}$. The phase space integral is evaluated as 
\begin{align}
\int \dd \Pi_{PS} = & 
\int \frac{\dd^3 \textbf{k}}{(2\pi)^3}\frac{1}{2E_N}\,
\int \frac{\dd^3 \textbf{p}}{(2\pi)^3}\frac{1}{2E_L}\,
\frac{2\pi}{2E_\Sigma}\delta(E_\Sigma-E_N-E_L) \notag \\
= & \int\limits_0^\infty \frac{\dd k\,k^2}{(2\pi)^3}\int \dd \Omega_k\frac{1}{2E_N}\,
\int\limits_0^\infty \frac{\dd p\, p^2}{(2\pi)^3}\int \dd \Omega_p\frac{1}{2E_L}\,
\frac{2\pi}{2 E_\Sigma}\delta(E_\Sigma-E_N-E_L) \notag \\
= & \frac{1}{32\pi^3}
\int\limits_0^\infty \frac{\dd k\,k^2}{E_N}\,
\int\limits_0^\infty \frac{\dd p\,p^2}{E_L}\,
\int\limits_{-1}^{1} \dd \cos\theta_{12} \frac{1}{E_\Sigma}\delta(E_\Sigma-E_N-E_L) \notag \\
= & \frac{1}{32\pi^3}
\int\limits_{m_{N}}^\infty \dd E_N\sqrt{E_N^2-m_{N}^2}\,
\int\limits_{M_L}^\infty \dd E_L\sqrt{E_L^2-M_L^2}\,
\int\limits_{-1}^{1} \dd \cos\theta_{12} \frac{1}{E_\Sigma}\, \notag \\
& \times 
\delta\left(\sqrt{q^2 + (m_\pm)^2}-\sqrt{k^2+m_{N}^2}-\sqrt{p^2+M_L^2}\right) \notag \\
\equiv & \frac{1}{32\pi^3}
\int\limits_{m_{N}}^\infty \dd E_N\sqrt{E_N^2-m_{N}^2}\,
\int\limits_{M_L}^\infty \dd E_L\sqrt{E_L^2-M_L^2}\,
\int\limits_{-1}^{1} \dd \cos\theta_{12} \frac{1}{E_\Sigma}\,
\delta\left(f(\cos\theta_{12}\right)),
\label{eq:phase_space_int_final}
\end{align}
where $\theta_{12}$ is the angle between the outgoing lepton and right-handed neutrino and

\begin{align}
f(\cos \theta_{12}) \equiv \sqrt{p^2+k^2+2pk\cos\theta_{12} + m_\pm^2}-\sqrt{k^2+m_{N}^2}-\sqrt{p^2+M_L^2}.
\end{align}

After a further evaluation we reach the expression  

\begin{align}
\int \dd \Pi_{PS} = \frac{1}{32\pi^3}
\int\limits_{m_{N}}^\infty \dd E_N\,
\int\limits_{E^-}^{E^+} \dd E_L\,
\int \dd \cos\theta_{12}\, \delta\left(\cos\theta_{12}-\cos\theta_{12}^\mathrm{min}\right),
\end{align}
with
\begin{align}
E^\pm \equiv & \pm \frac{\sqrt{\sqrt{E_N^2-m_{N}^2}\left( m_\pm^2 - (M_L + m_{N})^2\right)\left( m_\pm^2 - (M_L - m_{N})^2\right)}}{2 m_{N}^2}\notag \\
& \mp \frac{E_N\left(m_{N}^2+M_L^2-m_\pm^2\right)}{2 m_{N}^2},
\label{eq:Integration_boundaries}
\end{align}
and 
\begin{align}
\cos\theta_{12}^\mathrm{min} \equiv \frac{2E_LE_N + m_{N}^2 + M_L^2 - m_\pm^2}{2k\sqrt{E_N^2-m_{N}^2}}.
\label{eq:cos_minimum}
\end{align}
All reaction densities and washout terms include

\begin{align}
\frac{2}{32\pi^3} 
\int\limits_{M_N}^\infty \dd E_N\,
\int\limits_{E^-}^{E^+} \dd E_L \left(\frac{E_N}{2\sqrt{E_N^2 - M_N^2}} \pm \frac{1}{2}\right)
&\times 
\Bigg[2\sqrt{E_N^2-M_N^2}E_L \mp \left( 2E_N E_L + M_N^2 + M_L^2 - m_\pm^2 \right)\Bigg].
\label{eq:final_gamma_expression}
\end{align}

After inserting appropriate distribution functions we finally obtain
\begin{align}
\gamma^{LC}_{\alpha,ij} = & \frac{2}{32\pi^3} \int\limits_{m_{N}}^\infty\dd E_N\, \int\limits_{E^-}^{E^+} \dd E_L\,
\frac{1}{e^{E_N/T}+1}\,\left(\frac{1}{e^{E_L/T}+1}+\frac{1}{e^{(E_L+E_N)/T}-1}\right) \notag \\
\times &\left(\frac{E_N}{2\sqrt{E_N^2 - m_{N}^2}} + \frac{1}{2}\right) \left[2\sqrt{E_N^2-m_{N}^2}E_L - \left( 2E_N E_L + m_{N}^2 + M_L^2 - m_\pm^2 \right)\right] y_{\alpha i}^T \,y_{\alpha j}^\dagger.
\label{eq:gamma_LC}
\end{align}
\begin{align}
\gamma^{LV}_{\alpha,ij} = & \frac{2}{32\pi^3} \int\limits_{m_{N}}^\infty\dd E_N\, \int\limits_{E^-}^{E^+} \dd E_L\,
\frac{1}{e^{E_N/T}+1}\,\left(\frac{1}{e^{E_L/T}+1}+\frac{1}{e^{(E_L+E_N)/T}-1}\right) \notag \\
\times &\left(\frac{E_N}{2\sqrt{E_N^2 - m_{N}^2}} - \frac{1}{2}\right) \left[2\sqrt{E_N^2-m_{N}^2}E_L + \left( 2E_N E_L + m_{N}^2 + M_L^2 - m_\pm^2 \right)\right] y_{\alpha j}^T y_{\alpha i}^\dagger.
\label{eq:gamma_LV}
\end{align}
\begin{align}
\gamma^{LC}_{WQ,\,\alpha,ij} = & \frac{2}{32\pi^3} \int\limits_{m_{N}}^\infty\dd E_N\, \int\limits_{E^-}^{E^+} \dd E_L\,
\left(\frac{1}{e^{E_L/T}+1}\cdot\frac{1}{e^{(E_L+E_N)/T}-1}\right) \notag \\
\times &\left(\frac{E_N}{2\sqrt{E_N^2 - m_{N}^2}} + \frac{1}{2}\right) \left[2\sqrt{E_N^2-m_{N}^2}E_L - \left( 2E_N E_L + m_{N}^2 + M_L^2 - m_\pm^2 \right)\right] y_{\alpha i}^T y_{\alpha j}^\dagger.
\label{eq:gamma_LCWQ}
\end{align}
\begin{align}
\gamma^{LV}_{WQ,\,\alpha,ij} = & \frac{2}{32\pi^3} \int\limits_{m_{N}}^\infty\dd E_N\, \int\limits_{E^-}^{E^+} \dd E_L\,
\left(\frac{1}{e^{E_L/T}+1}\cdot\frac{1}{e^{(E_L+E_N)/T}-1}\right) \notag \\
\times &\left(\frac{E_N}{2\sqrt{E_N^2 - m_{N}^2}} - \frac{1}{2}\right) \left[2\sqrt{E_N^2-m_{N}^2}E_L + \left( 2E_N E_L + m_{N}^2 + M_L^2 - m_\pm^2 \right)\right] y_{\alpha j}^T y_{\alpha i}^\dagger.
\label{eq:gamma_LVWQ}
\end{align}
\begin{align}
\gamma^{LC}_{WC,\,\alpha,ij} = & \frac{2}{32\pi^3} \int\limits_{m_{N}}^\infty\dd E_N\, \int\limits_{E^-}^{E^+} \dd E_L\,
\left(\frac{1}{e^{E_N/T}+1}\cdot\frac{1}{e^{E_L/T}-1}\right) \notag \\
\times &\left(\frac{E_N}{2\sqrt{E_N^2 - m_{N}^2}} + \frac{1}{2}\right) \left[2\sqrt{E_N^2-m_{N}^2}E_L - \left( 2E_N E_L + m_{N}^2 + M_L^2 - m_\pm^2 \right)\right] y_{\alpha i}^T y_{\alpha j}^\dagger.
\label{eq:gamma_LCWC}
\end{align}
\begin{align}
\gamma^{LV}_{WC,\,\alpha,ij} = & \frac{2}{32\pi^3} \int\limits_{m_{N}}^\infty\dd E_N\, \int\limits_{E^-}^{E^+} \dd E_L\,
\left(\frac{1}{e^{E_N/T}+1}\cdot\frac{1}{e^{E_L/T}-1}\right) \notag \\
\times &\left(\frac{E_N}{2\sqrt{E_N^2 - m_{N}^2}} - \frac{1}{2}\right) \left[2\sqrt{E_N^2-m_{N}^2}E_L + \left( 2E_N E_L + m_{N}^2 + M_L^2 - m_\pm^2 \right)\right] y_{\alpha j}^T y_{\alpha i}^\dagger.
\label{eq:gamma_LVWC}
\end{align}

\bibliographystyle{JHEP}
\bibliography{keV_new_model}

\end{document}